\documentclass[vecphys]{svmult}

\newcommand{\tdecay}{{\text{decay}}}
\newcommand{\tviscous}{{\text{visc}}}
\newcommand{\tAlfven}{{\text{A}}}
\newcommand{\tOhm}{{\text{Ohm}}}

\newcommand{\ttdecay}{\tau_\tdecay}

\newcommand{\ttA}{\tau_\tAlfven}
\newcommand{\ttAO}{\tau_{\tAlfven,0}}
\newcommand{\ttOhm}{\tau_\tOhm}

\newcommand{\Alf}{Alfv\`en}
\usepackage{epsfig}
\usepackage{graphicx}
\usepackage{epic}
\usepackage{array}
\usepackage{curves}
\usepackage{amscd}
\usepackage{amsmath}
\usepackage{amssymb}
\usepackage{multirow}
\usepackage{natbib}
\usepackage{float}

\newcommand{\deriv}[3]{\frac{#3\hspace*{-.06em} {#1}}{#3\hspace*{.06em} {#2}}}

\newcommand{\parder}[2]{\deriv{#1}{#2}{\partial}}

\newcommand{\rezip}[1]{\frac{1}{#1}}

\newcommand{\grad}{\operatorname{grad}}
\newcommand{\curl}{\operatorname{curl}}
\newcommand{\dv}{\operatorname{div}}

\renewcommand{\exp}{\operatorname{exp}}

\newcommand{\ovec}[1]{{\mbox{\boldmath $#1$}}}
\newcommand{\rvec}{\ovec{r}}

\newcommand{\ervec}{\ovec{e}_r}

\newcommand{\ezvec}{\ovec{e}_z}
\newcommand{\zervec}{\ovec{0}}
\newcommand{\tA}{{\text{A}}}

\newcommand{\Bvec}{\ovec{B}}
\newcommand{\bvec}{\ovec{b}}

\newcommand{\uvec}{\ovec{u}}

\newcommand{\Pm}{\operatorname{\mathit{Pm}}}

\newcommand{\tmag}{{\text{mag}}}
\newcommand{\tkin}{{\text{kin}}}
\newcommand{\Emag}{E_\tmag}
\newcommand{\Ekin}{E_\tkin}

\newcommand{\myref}[1]{~\hspace{0pt plus 1pt minus 1pt}\ref{#1}}

\newcommand{\sectref}[1]{Sect.\myref{#1}}

\newcommand{\figref}[1]{Fig.\myref{#1}}
\newcommand{\tablref}[1]{Table\myref{#1}}
\newcounter{saveqn}
\newcommand{\alpheqn}{\refstepcounter{equation}\setcounter{saveqn}{\value{equati
on}}%
\setcounter{equation}{0}%
\renewcommand{\theequation}{\mbox{\arabic{chapter}.\arabic{saveqn}\alph{equation
}}}}
 
\newcommand{\reseteqn}{\setcounter{equation}{\value{saveqn}}%
\renewcommand{\theequation}{\arabic{chapter}.\arabic{equation}}}

\usepackage{makeidx}         
\usepackage{graphicx}        
\usepackage{psfig}           
\usepackage{multicol}        
\usepackage{url}             
\usepackage[bottom]{footmisc}

\makeindex             

\begin{document}

\title*{Turning Points in the Evolution of Isolated Neutron Stars' Magnetic Fields}
\titlerunning{Magnetic Fields of Isolated Neutron Stars}

\author{U. Geppert\inst{}}

\institute{
     Departament de F\'{\i}sica Aplicada, Univeritat d' Alacant, ap. Correus 99,
     03080 Alacant, Spain, \texttt{urme@ua.es}}

\maketitle

\begin{abstract}
During the life of isolated neutron stars (NSs) their magnetic field   passes through a variety of evolutionary phases. Depending on its strength and structure and on the physical state of the NS (e.g. cooling, rotation), the field looks qualitatively and quantitatively different after each of these phases. Three of them, the phase of MHD instabilities immediately after NS's birth, the phase of fallback which may take place hours to months after NS's birth, and the phase when strong temperature gradients may drive thermoelectric instabilities, are concentrated in a period lasting from the end of the proto--NS phase until 100, perhaps 1000 years, when the NS has become almost isothermal. The further evolution of the magnetic field proceeds in general inconspicuous since the star is in isolation. However, as soon as the product of Larmor frequency and electron relaxation time, the so--called magnetization parameter ($\omega_B \tau$), locally and/or temporally considerably exceeds unity, phases, also unstable ones, of dramatic changes of the field structure and magnitude can appear.\\
An overview is given about that field evolution phases, the outcome of which makes a qualitative decision regarding the further evolution of the magnetic field and its host NS. 

\end{abstract}

\section{Introduction}\label{sec:0}
The energy of the magnetic fields of neutron stars (NSs) is even for magnetar field strengths ($\sim 10^{15}$ G) negligible in comparison with the gravitational energy of even the least massive NSs. Except for the magnetars, this is true also when comparing with the rotational and thermal energy of the majority of NSs which appear as radio pulsars. Nevertheless, the magnetic field plays a decisive r{\^o}le for practically all observable quantities. Together with a sufficiently rapid rotation it allows them to detected as pulsars. For magnetars it is believed that their thermal radiation and burst activities are powered dominantly by their ultrastrong magnetic field (for a review see Harding \& Lai 2006). In standard NSs, possessing surface fields in the range of $10^{11.5 \ldots 13.5}$ G, observations as e.g. thermal radiation (see reviews of Page and Haberl in this volume), drifting subpulses (Gil et al. 2003), glitches (see, e.g., Ruderman 2005), cyclotron lines in X--ray spectra, or the variety of magnetospheric emissions (see, e.g., Becker \& Aschenbach 2002, Zavlin \& Pavlov 2004) are significantly affected by the magnetic field and consequences of its evolution. On the other hand, the cooling process affects the field evolution (see Yakovlev et al. 2001 for a core field; Page et al. 2000 for a crustal field), and the braking of the NS's rotation, almost completely determined by the dipolar magnetic field, may have a backreaction onto the field decay by affecting the flux expulsion from the suprafluid core (Alpar et al. 1984, Chau et al. 1992, Konenkov \& Geppert 2000 and references therein). The magnetic field evolution in millisecond pulsars will not be discussed here, since their progenitors went most likely through a phase of spin--up by accreting matter from a low mass companion star in a binary system (Bhattacharya \& van den Heuvel 1991, Urpin et al. 1998, Podsiadlowski et al. 2002), thereby manifolding the factors which influence the field evolution.\\
At first glance an isolated NS seems to be a sphere where its inborn magnetic field evolves only slowly because the electric conductivity is enormous and external influences don't affect it. However, there are periods during which the field evolution proceeds comparatively fast, sometimes even dramatic, and the field arrives repeatedly at a crossroad. There, depending on the physical state of the NS and on the strength and structure of the magnetic field itself, the field can evolve in two qualitatively different ways. Which of them is taken will have of course consequences for the further evolution of the whole NS. This is a natural implication of the tight connection between the magnetic and the thermal and rotational evolution.\\
Three of these phases appear quite early in a NS's life. The first turning point occurs immediately after a NS's birth. Here, the birth moment is understood to be that moment, when the proto--NS phase has been completed: the isolated NS has almost reached its final mass--radius relation and density profile, convective motions have ceased and the matter is uniformly rotating, and in a liquid and normal (i.e. non--suprafluid) state. At this stage the field, whatever structure and strength it has acquired after collapse and proto--NS stage, must possibly  go through magnetohydrodynamic (MHD) instabilities and may suffer from the most dramatic changes conceivable for any magnetic field evolution: during the first $\lesssim 10$ seconds of its life MHD instabilities may reduce an initially perhaps $\sim 10^{15}$ G dipolar field down to standard pulsar field strengths of $\sim 10^{12}$ G, depending on the rotation period and the inclination angle between magnetic and rotation axis (Geppert \& Rheinhardt 2006).\\
The second point comes when after the supernova explosion the hypercritical fallback accretion, produced by a reversed shock, reaches the NS surface. In the case of SN 1987 A this happened about $2$ hours after bounce (Chevalier 1989). However, depending on the detailed properties of each particular supernova this moment will vary (Colpi et al. 1996). In some cases and influenced by many factors depending on the medium surrounding the NS, its kick and rotation velocities, and on its magnetic field strength just after the mentioned MHD instabilities, this fallback can be significantly reduced or will not take place at all. When it, however, proceeds similar as  estimated for SN 1987 A, it will submerge a pre--existing magnetic field in the crust, perhaps down to the crust--core interface. The strength of the fallback accretion and the conductive properties of the crust decide whether the NS appears as a radiopulsar already at its birth or, in case of weak fallback and shallow submergence, after say some $100$ years (Muslimov \& Page 1995). For heavier fallback and deeper submergence the rediffusion of the field to the surface takes millions or hundreds of millions of years and the NS is either radio quiet or its active age (given by the rotational period and its time derivative, $P$ and $\dot{P}$), is much larger than its real age, making relatively young NSs to old looking ones (Geppert et al. 1999).\\
The third turning point happens when the conditions are such that a thermoelectric instability may efficiently convert thermal into magnetic energy (see Urpin et al. 1986, Geppert \& Wiebicke 1991 and references therein). The preferred region where that instability takes place is the low density liquid shell ($\rho \le 10^{10}$g cm$^{-3}$) where a sufficiently strong temperature gradient prevails. The thermoelectric instability in that layer will act as long as $T_{s6}^4/g_{s14} > 100$, where $T_{s6}$ is the surface temperature in $10^6$ K and $g_{s14}$ is the gravitational acceleration at the NS's surface in $10^{14}$ cm s$^{-2}$ (Wiebicke \& Geppert 1996). For a typical NS with $g_{s14} \approx 1$ this means that the surface temperature should exceed $3\cdot 10^6$ K, which, depending on the cooling scenario is likely only for the first $100$ years, in case of the slowest cooling perhaps $1000$ years of NS's life. During the initial, linear stage of the thermoelectric instability only small scale ($\sim 100$ m) toroidal field components are exponentially growing on a timescale of $50\ldots 100$ days.  After about $10$ years nonlinear interactions raise to large scale toroidal components. However, the studies stopped at that point and all attempts failed up to now to create the observable poloidal large scale fields of $10^{12\ldots 13}$ G at the surface of the star by that instability.\\
There are however very promising regions in young NSs, where the thermoelectric instability may act even more efficiently than in the outer liquid crust. As shown by Gnedin et al. (2001) there exist extremely strong temperature gradients deep in the crust during first year of the early thermal relaxation phase. A strong magnetic field created in those layers by the thermoelectric instability during the first hours of the NS's life could alter the magneto--thermal evolution qualitatively.\\
The fourth "turning point" of the magnetic field evolution in an isolated NS can not be dedicated to a certain phase of its life but appears always if the magnetization parameter $\omega_B \tau$ significantly exceeds unity, either temporally or spatially. Here, $\omega_B = e B/(m_e^{\ast} c)$ is the Larmor frequency, where $B$ is the strength of the magnetic field, $e$ the elementary charge, $c$ the velocity of light and $m_e^{\ast}$ the effectice electron mass which depends on density and chemical composition. The relaxation time $\tau$, that is the period between collisions of the electrons with the relevant impact partners as, e.g., ions, phonons or impurities in the crust or protons and neutrons in the core, is a complicated function of the density, temperature, chemical composition, impurity concentration, the occurence of superfluidity, and perhaps other factors and increases usually significantly when the NS cools down. The magnetization parameter may therefore exceed unity in different phases of a NS's life and perhaps only locally in certain regions of the NS: when either the magnetic field strength or the relaxation time increases beyond a certain threshold. Thereby, the strength of one quantity may overcompensate the smallness of the other. In crusts of magnetars, e.g., the relaxation time is relatively short because of the high temperatures, but the extremely strong magnetic field ensures that $\omega_B \tau \gg 1$ there.\\
Two consequences of situations with $\omega_B \tau \gg 1$ will be discussed in this review. 
Firstly, the situation will be considered when a large magnetization parameter causes a domination of the magnetic field evolution by the Hall--drift. Although the Hall--e.m.f. is conservative, it redistributes magnetic energy from large scales into smaller ones which eventually decay much faster ohmically than pure dipole fields would do. Goldreich \& Reisenegger (1992) studied this process in detail, coining for the continuous transfer of magnetic energy towards smaller scaled field modes the term "Hall--cascade".\\
Alternatively to or simultaneously with the Hall--cascade, a non--local unstable energy transfer may proceed from a relatively large scaled background field into smaller scaled perturbations, if the second derivative of the background field with respect to at least one spatial variable is non--zero, i.e. if this field is sufficiently curved (Rheinhardt \& Geppert 2002). This instability is driven by the shear in the electron velocity of the current maintaining the background field (Cumming et al. 2004, Rheinhardt \& Geppert 2005 and references therein). The large braking indices observed in some middle aged pulsars could be explained very well by an energy drain out of the dipolar field, which determines the braking, into smaller scaled perturbations just on the time scale of the Hall instability (Geppert \& Rheinhardt 2002). In all, both the Hall--cascade and the Hall--instability will play a role  for the transient acceleration of the magnetic field decay as well as for the generation of strong small scale field structures at the surface, necessary for the pulsar mechanism to work. Their relative importance, however, has not yet been clarified.\\
Secondly, the situation will be considered when a large magnetization parameter causes significant deviations from the isotropic heat transport through the NS's crust, leading to comparatively small hot spots around the magnetic poles, leaving the largest part of the surface so cool that it can not contribute remarkably to the observed X--ray flux of thermally emitting isolated NSs. Meanwhile a large number of X--ray and combined X--ray and optical observations (revealing in some cases a significant larger optical flux than to be expected by the continuation of the blackbody spectrum  from the X--ray energies, called "optical excess") of these NSs (called "The Magnificent Seven") are available and both spectral and lightcurve analyses justify quite well the conclusion, that the surface temperatures are indeed highly anisotropic (see e.g. Becker \& Tr{\"u}mper 1997, Pons et al. 2002, Burwitz et al. 2003, Tr{\"u}mper et al. 2004, Haberl 2004, and Haberl 2005; see also the review of Haberl in this volume). It is known  already since Greenstein \& Hartke (1983) that a dipolar magnetic field which penetrates the envelope (upper crustal region where $\rho < 10^{10}$ g cm$^{-3}$) being stronger than, say, $10^{11}$ G causes a surface temperature gradient from the pole to the equator. The observed surface temperature profiles, however, can not be explained by assuming anisotropy of the heat transfer in the envelope only, but considering the whole crust from the crust--core interface up to the bottom of the envelope as isothermal. A satisfactory agreement with the observational facts can be reached, when the anisotropy of the heat transport in the whole crust is taken into account (see Geppert et al. 2004 and 2006, P\'{e}rez-Azor\'{\i}n et al. 2005 and 2006, Zane \& Turolla 2005 and 2006). The strong crustal field enforces heat transport prevailingly parallel to the field lines, while the heat flux perpendicular to them is strongly suppressed. In contrast, a star centered core field causes only very little anisotropy of the heat flux through the crust. Only a crustal field consisting of sufficiently strong poloidal and toroidal components may create such a temperature distribution at the bottom of the envelope that the further heat transfer through it creates the observed large surface temperature gradients in meridional direction. Although due to the variety of possible emission processes at the NS surface, being either an atmosphere, a liquid or a solid, having an unknown chemical composition etc., other reasons for the observed temperature profiles and the "optical excess" are conceivable (see e.g. van Adelsberg et al. 2005), the good agreement of the observations with the model calculations of the anisotropic heat transfer is a quite reliable hint for the existence of strong crustal field configurations.\\
As can be seen already from the references, the instabilities immediately after NS's birth and the effect of a large magnetization parameter on field evolution and heat transport currently gain intensive attention, while the consideration of the consequences of the supernova fallback as well as the study of thermoelectric instabilities in NSs is recently not in the very focus of the community. I believe, however, that all these five possible "turning moments" of the field evolution deserve further research activities and I will now try to describe the corresponding physics and the results available up to now in some detail.\\

\section{MHD Instabilities Immediately After Birth: Magnetar or Radiopulsar?} \label{sec:1}

\begin{figure}
\centerline{\hspace*{-0.1cm}\epsfig{file=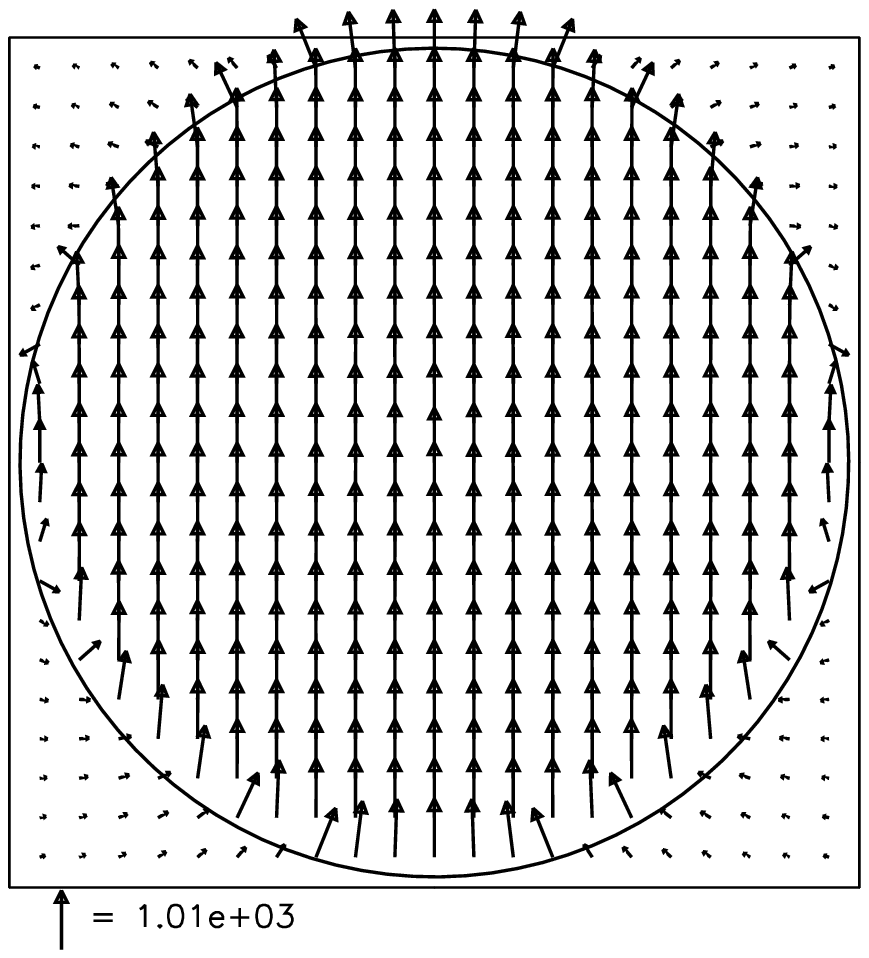,width=9cm,clip=}\hspace*{-2cm}\epsfig{file=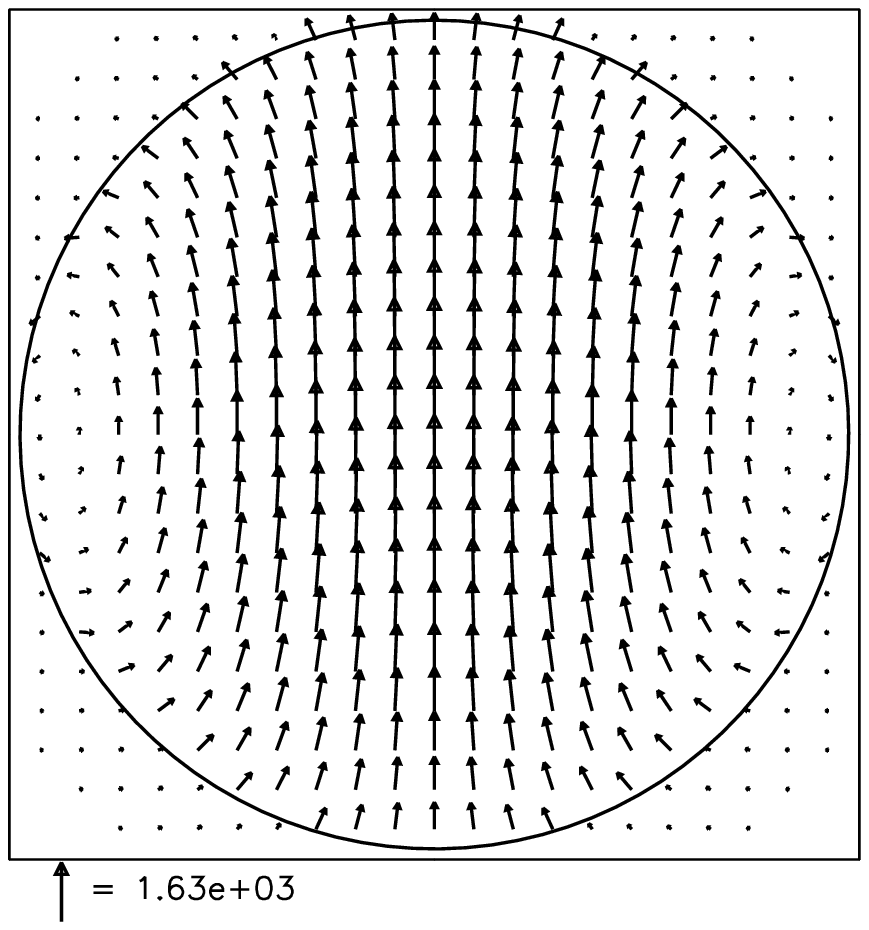,width=9cm,clip=} } 
\caption{Left panel: internal uniform field according to \eqref{uniform} in a meridional plane. The arrow indicated below marks a magnetic field of $10^{15}$G. Right panel: dipolar magnetostatic equilibrium field according to \eqref{poleq}. The arrow indicated below marks a magnetic field of $1.63\cdot10^{15}$G. }
\label{fig:initial_field}
\end{figure}

\begin{figure}
\centerline{\epsfig{file=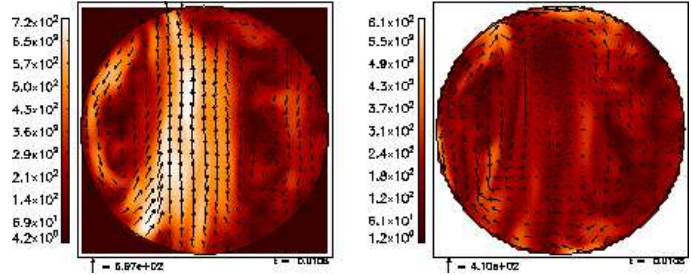,width=9cm,clip=} }
\caption{Field geometries of magnetic field (left) and flow (right) around the end of the dramatic field reduction phase (see upper panels of {\figref{fig:mageq_instab}} at $t\approx 0.01$) in  a meridional plane for the dipolar equilibrium model with $q_P$=12, $\Pm=1$, $\alpha=0$. Arrows indicate vector components parallel to the paper plane. Their maxima are $6.93\cdot 10 ^{14}$ G and $8.3\cdot10^{6}$ cm s$^{-1}$, respectively. Colors encode the field moduli: the brightest tone corresponds to $7.2\cdot 10^{14}$ G  and  $1.2\cdot10^{7}$ cm s$^{-1}$, respectively.}
\label{fig:mageq_transient}
\end{figure}

When the proto--NS phase is completed a (quasi--)isolated NS is born. Unfortunately, no newborn NS is observable. The nearly simultaneously formed supernova remnant obscures the star for hundreds of years almost completely. Therefore, theoretical models have to promote the understanding of the physical conditions at a NS's birth.\\
During the pre--supernova phase of the massive progenitor star, during its core's collapse and during the proto--NS stage, which lasts about $30$ seconds, any pre--existing magnetic field can be amplified tremendously.\\
In the pre--supernova phase the so--called "Spruit dynamo" (Spruit 1999 and 2002) is probably a powerful amplification mechanism. It transfers the energy of differential rotation into magnetic energy. A poloidal seed field will be wound up by differential rotation, thereby creating a toroidal field component. This component undergoes a MHD instability (Tayler (1973)), thereby forming poloidal components which can tap the energy of differential rotation again, closing in that way the dynamo cycle. Estimates show, that the toroidal component of the generated field may reach up to $5\cdot 10^9$ G for a $15 M_{\odot}$ star while its poloidal component  will be considerably weaker (Heger et al. 2005).\\
In the core collapse another enhancement of the magnetic field strengh proceeds, at least simply by flux conservation in the highly conductive matter. Depending on the mass of the progenitor, the amplification factor can be as large as $10^4\ldots 10^5$.\\
The proto--NS phase is characterized by vigorous convective motions which are driven by the usual Ledoux instability and/or by doubly diffusive instabilities (Keil et al. 1996, Pons et a. 1999, Miralles et al. 2000,  2002, and  2004, Bruenn et al. 2004). Thompson \& Duncan (1993) argue that under the conditions prevalent in the proto--NS these rapid convective motion  may generate fields as strong as $10^{16}$ G by dynamo action. However, the situation in the proto--NS is characterized by extremely large magnetic Reynolds numbers $R_m \sim 10^{19}$ which are by about $16$ orders of magnitude overcritical with respect to the onset of the dynamo. Since the proto--NS convection is anyway a transient phenomenon, lasting at most $\sim 20$ seconds, there exist various scenarios of magnetic field amplification (see Rheinhardt \& Geppert 2005a), including that the dynamo acts only at the end of the convective stage, when the convective velocity decreases. Then, $R_m \lesssim 1000$ and the backreaction of the growing field can not longer efficiently counteract the dynamo process.\\
Since the fluid in the proto--NS is - at least during its early stage - for sure in turbulent motion and differentially rotating, the description of the magnetic field evolution by mean field models, i.e. by $\alpha$-- and/or $\Omega$--dynamos, is suggestive (see Bonanno et al. 2006 and references therein). However, the mean-field coefficients of these models are derived in second order correlation approximation (SOCA). The SOCA results (in the high-conductivity limit) have been justified so far for a Strouhal number $St = \tau V/l << 1$ and are perhaps still useful as ''order of magnitude estimates'' if $St \sim 1$ (see Krause \& R{\"a}dler (1980) Eqs. 3.12 and 4.10, Petrovay \& Zsargo (1998)).  In the proto--NS, however, the minimum of both the overturn time  and the characteristic life time of velocity pattern $\tau \gtrsim 10$ ms, the convective velocity $V \sim 10^8 \ldots 10^9$ cm s$^{-1}$ and the typical scale of the convective eddies $l \lesssim 10^5$ cm; therfore $St\gtrsim 10$, making the results obtained by Bonanno et al. (2006) questionable.\\
Another field amplifying instability, the magnetorotational instability (MRI), acts very efficiently during the first $\lesssim 1$ second after bounce. The MRI is supposed to be able to tap the energy of differential rotation for the generation of fields exceeding even $10^{16}$ G (Akiyama et al. 2003, Thompson et al. 2005).\\
Therefore, though a satisfactory understanding of the field amplification processes during pre--NS--stages is still not well--elaborated, there are good reasons to assume that at the beginning of its life the NS is endowed with an ultrastrong magnetic field, perhaps in excess of $10^{15}$ G.\\
At the same time, the NS is likely in a state of very rapid ($P < 60$ ms) but rigid rotation.  Ott et al. (2006) argue that NSs reach at the end of the proto--NS phase the state of rigid rotation. The growth of the magnetic field in that phase is initially at least partly powered by the differential rotation. However, the growing magnetic field causes via the MRI a turbulent viscosity, much larger than the molecular one, which enables a rapid angular momentum redistribution. Therefore, at least at the end of the proto--NS phase, when convection ceased and the field is as strong as $\sim 10^{15}$ G, the state of uniform rotation will be reached, which is the lowest--energy state of a rotating body for a given angular momentum.\\
Another statement about the newborn NS can be made with great certainity: since the temperature at birth exceeds $10^{10}$ K, the complete NS is liquid, the crystallization of the crust has not yet been started and the core matter has not yet performed the transition into the superfluid state.\\
It is observationally evident that a NS which starts its life with a magnetic field exceeding $10^{15}$ G will evolve in a completely different manner than another which starts with a typical radio pulsar field of $10^{11\ldots 13}$G or, practically non--magnetized, as a millisecond pulsar with $\sim 10^8$ G. Therefore, the knowledge about the initial magnetic field strength and structure is basic for the understanding of the further evolution of NSs. More precisely, the question arises whether the inborn field can reach in a relatively short time - until the onset of crystallization and superfluidity - a stable equilibrium configuration. This is the old question for stable magnetostatic equilibria in a conducting sphere.\\
Prendergast (1956) already discussed that the coexistence of a poloidal and toroidal magnetic field can enable an equilibrium configuration within an infinitely conducting sphere. It was Wright (1973), who postulated the stability of a magnetic field configuration the poloidal and toroidal constituents of which are of comparable strength. Tayler (1973), Markey \& Tayler (1973), Wright (1973), and Pitts \& Tayler (1985) performed a series of analytical stability analyses of magnetic field configurations in stars. The outcome of these analyses is that a purely toroidal axisymmetric field without rotation in a stratified medium is always unstable with respect to large scale perturbations. The growth rate of this instability  is in the order of the \Alf{} frequency $\Omega_A= v_A/r$, where $v_A = B/\sqrt{4\pi\rho}$, $B$ the magnetic field strength,  $\rho$ the density, and $r$ the radial coordinate. This ``Tayler--instability'' acts locally in meridional ($r,\theta$) planes but globally in the azimuthal ($\varphi$) direction.\\
Markey \& Tayler (1973) and Wright (1973) performed similar adiabatic stability analyses for axisymmetric poloidal fields and concluded its general instability if at least some of its field lines are closed within the star. The instability generates small scale, rapidly decaying field structures on the same time scale as for a toroidal field. They discuss the possibility that sufficiently rapid rotation and/or the simultaneous existence of equipollent  toroidal and poloidal components may stabilize the axisymmetric background fields. Pitts \& Tayler (1985) performed an analytical stability analysis of a very special combination of toroidal and poloidal fields also with respect to rotation and to the magnetic inclination angle $\alpha$. They found a tendency of rotation to counteract the instability but concluded that "rotation is unlikely to lead to a complete stability of general magnetic field configurations".\\
In 2005 the interest in the stability of magnetostatic equilibria in stars became revitalized by a series of papers: Braithwaite \& Spruit (2004)and (2006), Braithwaite (2005), Braithwaite \& Nordlund (2006). The latter authors  studied the stability of magnetic fields in the radiative interiors of non--rotating Ap stars and found that  any random field is generally unstable but evolves in a stably stratified star towards a ``twisted torus'' configuration with approximately equipollent toroidal and poloidal components.
Braithwaite \& Spruit (2006) considered the stability of MHD configurations in magnetars. They found that, quite similar to the Ap stars, stable magnetic fields exist when being concentrated to a relatively small region around the center of the star and after having evolved into the poloidal--toroidal twisted torus shape.\\
While Braithwaite (2006) showed that sufficiently fast rotation may stabilize purely toroidal field configurations, until now the effect of rotation on the stability of purely poloidal field configurations has not been considered in general.\\
When NSs have inborn magnetic fields in the order of $10^{15}$ G, than the corresponding \Alf{} crossing time $\ttA= r/v_A$ is very short; for $r = R = 10^6$ cm and $\rho = 2\cdot 10^{14}$ g cm$^{-3} \approx 50$ ms. The rotation period at the NS's birth however, is supposed to be even smaller, at least in some cases, theoretically it can be as small as $\sim 0.8$ ms (for a $1.4 M_{\odot}, R = 10^6$ cm NS, Villain et al. 2004). Therefore, rotation in new--born NSs and its effect on any inborn magnetic field can for sure not be neglected. Since the dipolar magnetic field determines both observability and rotational evolution, Geppert \& Rheinhardt (2006) studied the consequences of rotation for the stability of such fields in detail.\\
Besides the above mentioned assumptions, that the whole star is liquid and in a normal state (for at least the first few $100$ seconds), that it rotates rigidly, and that it has an inborn field in the order of $10^{15}$ G, they assumed incompressibility, constant density, maintenance of the spherical shape of the star, and uniform rotation. The assumption of incompressibility can be  quite well justified {\it{ a pasteriori}} by comparing the maximum fluid velocities with the sound velocity. The deviations from a spherical shape can be neglected as long as both the magnetic and kinetic energies inherent in the conducting sphere are small in comparison with its gravitational binding energy. However, the assumption of constant density for a NS is of course a questionable one, the more since a stable stratification exerts a stabilizing effect on to the field. Although a certain amount of compressibility gives rise to other (e.g. Parker) instabilities, and although the bulk of the induced flow is in the core where the density gradient is by far not as large as in the crustal region, further calculations have to relax that premise.\\
The preceding considerations result in the following ruling equations for magnetic field $\Bvec$ and velocity $\uvec$ and corresponding boundary conditions  written in dimensionless form:
\begin{equation}
\begin{alignedat}{2}
\parder{\Bvec}{t} &= \Delta\,\Bvec + \curl(\uvec\times\Bvec) &\qquad\text{for} \quad r \le 1 \\
\curl \Bvec &= \zervec  &\qquad\text{for} \quad r > 1 \\
\dv\Bvec &= 0   &\qquad\forall r                                                      \\
[\Bvec] &= \zervec &\qquad\text{for}\quad r=1
\end{alignedat}\label{ind}
\end{equation}
\begin{align}
&\hspace*{-4mm} \left.\begin{alignedat}{3} \hspace*{-1.5cm}\parder{\uvec}{t} &=  &-&\,\nabla( p - \rezip{2}\Omega^2 r^2 \sin^2\vartheta+\Phi ) \\
                                                                                  &  &- &\,(\uvec\nabla)\uvec  + \Pm\Delta\,\uvec \\
	                                                                          &  &- &\,2 \Omega \ezvec\times\uvec + \curl\Bvec\times\Bvec\\ 
\dv\uvec&= \;0   
\end{alignedat}\,\right\}\;\; \text{for}\quad r \le 1\hspace{-3mm} \label{mom}\\*[0.5mm]
&\hspace*{1cm}  \left.\begin{aligned}
u_r &= 0 \\
(\hat{D}(\uvec)\cdot\ervec)_{\vartheta,\varphi}& = 0 
\end{aligned}\;\right\}  \qquad\text{for} \quad r=1\hspace*{-3cm}\label{sfRB}
\end{align}
Here, length is normalized on the NS radius $R$, time on the magnetic diffusion time
$\ttOhm=R^2/\eta$, where $\eta=c^2/4\pi\sigma$  is the magnetic diffusivity and $\sigma$ the electric conductivity. $[.]$ denotes the jump of a quantity across a surface. The magnetic field is measured in units of $ (4\pi\rho)^{1/2}\eta/R$, the velocity in units of $\eta/R$, the pressure $p$ in units of $\rho\eta^2/R^2$, and the gravitational potential $\Phi$ in units of $\eta^2/R^2$. $\hat{D}(\uvec)$ denotes the deformation tensor. $(r,\vartheta, \varphi)$ are spherical co--ordinates, the polar axis ($\vartheta=0$, $\upuparrows \ezvec$) of which coincides with 
the axis of rotation.\\
Equations \eqref{ind} and \eqref{mom} contain two parameters which together with the initial field strength $B_0$ (for a fixed initial field geometry) define the problem 
completely: the magnetic Prandtl number $\Pm$ and the normalized rotation rate $\Omega$. $\Pm$, is the ratio of kinematic viscosity, $\nu$, and magnetic diffusivity, $\eta$, and is chosen to be 0.1, 1, and 10, resp., what is partly dictated by numerical restrictions, but on the other hand represents a subset of the NS--relevant range determined by the prevailing densities ($\rho= (1\ldots2.8)\cdot10^{14}$ gcm$^{-3}$) and temperatures ($\sim 10^{11\ldots9} $K). 
These values result in $\Pm =0.1 \ldots 6\cdot10^8$  (Cutler \& Lindblom 1987, Yakovlev \& Shalybkov 1991) reflecting mainly the strong temperature  dependence of $\Pm\propto T^{-4}$.\\ 
The two remaining parameters are best expressed as ratios of  characteristic times: $\Omega$ by $q_P=P/\ttAO$, and $P$ the rotation period of the NS, where $\ttAO= R/v_{\tA,0}$ is the \Alf{} crossing  time related to the initial field, and the initial field amplitude $B_0$ by $q_{B_0}=\ttAO/\min\{\ttdecay^B,\ttdecay^u\}$,.
Note, that the ``decay times'' of the magnetic and velocity fields, $\ttdecay^B$, $\ttdecay^u$, are different from the
often used magnetic and viscous ``diffusion times'', $\ttOhm=R^2/\eta, \; \tau_\tviscous =  R^2/\nu$, by factors $1/\pi^2$ and $0.06676$, respectively.\\
For $q_{B_0}$  values $\sim 10^{-2}$ are assumed, which are of course by far too large for highly magnetized young NSs where this ratio can go down to $10^{-16}$ for a surface magnetic field of $10^{15}$ G. However, as long as the growth/decay times of the examined perturbations are in the order of at most a few \Alf{} times, dissipation of the background state does not  affect their linear stage (quasi--stationary approximation). The artificially enhanced dissipation is a concession to numerical feasibility only, in order to avoid excessive requirements 
for spatial resolution (cf. Braithwaite \& Spruit 2004, where $q_{B_0}$ was chosen to be 0.1). Hence, the use of the results from the nonlinear stage has to be considered with caution when the elapsed time since birth exceeds about $100 \ttAO$. However, the non--linear treatment returns at least reliable satuartion values.\\
For the models presented here the following parameters were choosen:
\begin{equation}
R=10^6\text{cm}\, ,\:
\rho=2\cdot10^{14}\text{g/cm}^3\, ,\:
\tilde{B}(t=0)=10^{15} \text{G},
\label{denorm}
\end{equation}
where $\tilde{B}$ is the de--normalized r.m.s. value. Therefore, $\ttAO=0.05$ s. With rotation periods  chosen between 0.6 ms and 0.6 s,  where the former value is somewhat below the generally accepted minimum  for new--born NSs, but the latter is in that respect not very likely, these model parameters result in $q_P=0.012\ldots12$. As a reference and for comparison with known results for non--rotating stars $q_P=\infty$ is considered, too.\\
Two qualitatively different initial field configurations are considered. Each of these consists of an axisymmetric dipolar background field, the stability of which is examined, and imprinted magnetic perturbations. The fluid is assumed to be at rest initially, and the initial magnetic inclination $\alpha$ is defined by the dipole axis. 
The two background fields are shown in~\figref{fig:initial_field} for $\alpha=0$. The energy of the perturbations is set to 0.1\% of the magnetic background energy, and their geometry ensures that the initial state has no preferred equatorial symmetry. 
The uniform background field configuration, a more academic example, consisting of a field uniform throughout the sphere, but being a dipolar vacuum field outside: 
 \begin{equation}
\begin{alignedat}{2}
\Bvec &= \phantom{-}B_0\ezvec &\quad \text{for}\; r\le 1\\
\Bvec &= - B_0 \curl\rvec\times\nabla\left (\frac{\cos\vartheta}{2r^2}\right)&\quad \text{for}\; r > 1\\
[B_r] &= 0  &\quad \text{for}\; r= 1
\end{alignedat}\label{uniform}
\end{equation}
Here, $B_0$ denotes the polar surface field strength. Of course, the continuity of the normal component of $\Bvec$ has to be required, but the tangential components remain discontinuous and give rise to surface currents. This model field was chosen, because it was the one considerations on magnetic stability in NSs were first exemplified on by Flowers \& Ruderman (1977) and because Braithwaite \& Spruit (2006) report on its instability in the non--rotating case.\\
A more realistic initial configuration consists of the poloidal magnetostatic equilibrium field the angular dependence of which is the same as for a dipolar 
field (Roberts 1981, Rheinhardt et al. 2004):   
\begin{equation}
\begin{alignedat}{2}
\Bvec &= \phantom{-}B_0\;\curl\rvec\times\nabla\left (\frac{1}{4}(3r^3-5r) \cos\vartheta\right)&\quad \text{for}\; r\le 1\\
\Bvec &= - B_0\; \curl\rvec\times\nabla\left (\frac{ \cos\vartheta}{2r^2}\right)&\quad \text{for}\; r > 1\\
[\Bvec] &= 0 & \quad \text{for}\; r= 1
 \end{alignedat}\label{poleq}
 \end{equation}
As the Lorentz force of this field is a pure gradient it can, in the constant--density case, be balanced by the quantity $-\nabla\big( p - (\Omega^2 r^2 \sin^2 \vartheta)/2 + \Phi \big)$ (see Eq. \eqref{mom}). If stable, this state could be a final equilibrium to which an arbitrary initial configuration relaxes during the fluid stage of a newly born NS. As known from analytical studies of Markey \& Tayler (1973), however, any purely poloidal field is in the absence of rotation surely unstable, hence the evolution of the field \eqref{poleq} under the influence of rotation is a crucial question.\\
Because the star is modeled as a  spherical body and incompressibility is assumed, the problem is especially suited to be tackled by a spectral method. The one used here employs an expansion of $\Bvec$ and $\uvec$ into their modes of free decay.\\
The numerical solution of Eqs.~\eqref{ind},~\eqref{mom} returns qualitatively very different results for the magnetic field evolution in dependence on the rotational period and the inclination angle $\alpha$. If the rotation is fast enough and/or the inclination angle sufficiently small, the background field is stable and reaches quite fast a stable equilibrium configuration. If, in contrast, rotation is too slow and/or $\alpha$ too large, the background field becomes unstable and looses almost all of its initial energy by transfering it to small scale modes of the velocity and the magnetic field, for which dissipation acts efficiently. For the parameter range considered up to now, the transition from stabilization to unstable behaviour happens at a rotational period $P\gtrsim 6$ ms and $\alpha \approx 45^\circ$.
To illustrate it, in ~\tablref{tab:res_mageq} the results for a model with $\Pm = 1$ and the two different initial background field configurations are presented.
\begin{table*}
\caption{Left part: results for the dipolar magnetostatic equilibrium model. All quantities are taken after a period of $\ttdecay^B$. Subscripts ``kin'' and ``mag'' refer to velocity and magnetic--field related quantities, respectively. The calculations for the non--rotating NS, $q_P=\infty$, were performed only for $\alpha=0$ because in this case the choice of the axis is of course arbitrary. Right part: the corresponding results for the internal uniform field model.} 
\label{tab:res_mageq}
   \begin{minipage}[t]{0.5 \linewidth}
    \begin{center}
      \begin{tabular}{cccc}
      \hline
      \hline\\*[-2.5mm] 
$\alpha\, (^\circ)$\hspace{0.0cm}  & $q_P$  \hspace{0cm}&  $ \Emag/\Emag^\tOhm$\hspace{0.0cm} & $ \Ekin/\Emag$\\*[1mm]
       \hline\\*[-2.5mm]
\multirow{4}{8mm}{ 0}  &  $\infty$ &  0.0002 & 4.67     \\
			& 12.       &  0.0004 & 6.6      \\
			& 1.2       &  0.0076 & 0.0007   \\
		        & 0.12      &  0.98   & 0.00003  \\*[.5mm]
                        		    \hline \\*[-2.5mm]
\multirow{2}{8mm}{ 45 }& 1.2       &  0.075  & 0.001    \\
		        & 0.12      &  0.824  & 0.00005  \\*[.5mm]
                        		    \hline \\*[-2.5mm]
\multirow{4}{8mm}{ 90 }& 12.       &	0.0033 & 0.64    \\
			& 1.2       &  0.043  & 0.003    \\
		        & 0.12      &  0.14   & 0.00014  \\
                       & 0.012     &	0.98   & 0.0013  \\*[.5mm]
                                           \hline\\*[-2.5mm]
    \end{tabular}
   \end{center}
  \end{minipage}%
\begin{minipage}[t]{0.5 \linewidth}

\vspace{-2.75cm}
\begin{center}
      \begin{tabular}{cccc}
         \hline
         \hline\\*[-2.5mm]     
$\alpha\, (^\circ)$\hspace{0.0cm} & $q_P$  \hspace{0cm} &  $\Emag/\Emag^\tOhm$\hspace{0.0cm} & $\Ekin/\Emag$ \\*[1mm]
       \hline\\*[-2.5mm]
\multirow{4}{8mm}{ 0}  &  $\infty$  & 0.00009 &  147.6  \\
                       & 12.      & 0.001   &  2.92 \\
                       & 1.2      & 0.083   &  0.0034\\
                       & 0.12     & 1.02    &  0.00002\\*[.5mm]
                        		    \hline\\*[-2.5mm]
\multirow{2}{8mm}{ 45 }& 0.12     & 0.56    &  0.00006\\
                       & 0.012    & 1.006   &  0.0023  \\*[.5mm]
                        		    \hline\\*[-2.5mm]
\multirow{4}{8mm}{ 90 }                                  \\
                                                          \\
                       &  0.12    & 0.193   &  0.0005 \\
                       &  0.012   & 0.976   &  0.0013  \\*[0.5mm]
                        		    \hline
      \end{tabular}
    \end{center}
  \end{minipage}
\end{table*}
Note, that for the model parameters~\eqref{denorm} $q_P=0.12$ corresponds to $P=6$ ms. Most informative about the effect of rotation and inclination is the third row, where the final magnetic energies after equal evolution times is related to the final energy of the purely ohmic case $\Emag^\tOhm$, i.e. when no coupling to the fluid flow is allowed. It is clearly seen that with increasing rotational velocity, an increasingly smaller part of the magnetic energy is dissipated and/or transferred into kinetic energy (fourth row). That an increasing $\alpha$ exacerbates the stabilization of the background field is also shown by the fact, that for $\alpha= 90^\circ$ an extremely fast rotation with $P=0.6$ ms would be necessary to stabilize the dipolar magnetostatic equilibrium field. It is informative to consider in comparison the evolution of the initial uniform background field (left panel of ~\figref{fig:initial_field}). The general tendencies are the same as for the dipolar equilibrium field. Again, the significant influence of $\alpha$ on the stability is proven.\\
It is interesting to see how an unstable background field configuration evolves through the instability. In ~\figref{fig:mageq_transient} a snapshoot of both the velocity and magnetic field evolution  is shown at $ 10\ttA = 0.5$ s for a newborn NS rotating initially with $P=60$ ms (second line in ~\tablref{tab:res_mageq}).

\begin{figure}
\centerline{\epsfig{file=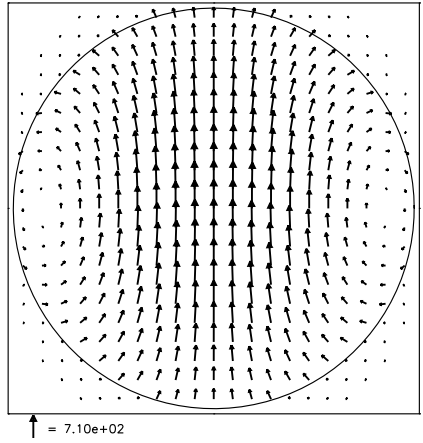,width=9cm,clip=} } 
\caption{Final field geometry for the internal uniform field model in a meridional plane (the field is almost exactly axisymmetric) for $\alpha=0$, $q_P=0.12$, $\Pm=1$. With the denormalization based on Eq. \eqref{denorm} the maximum field  strength is $\approx 7.1\cdot10^{14}$ G.}
\label{fig:pol_final}
\end{figure}

A comparison of both the final spectra and field geometries of the initially quite different background field configurations shown in ~\figref{fig:initial_field} for $P=6$ ms and $\alpha=0^\circ$ (fourth lines in ~\tablref{tab:res_mageq} gives strong evidence for a tendency to approach the same state for $t\rightarrow\infty$ (see ~\figref{fig:pol_final}). 
Not only the difference in the initial geometry (cf. ~\figref{fig:initial_field}), but also the difference in the initial energies is obviously equalized after having gone through the nonlinear stage. The relative r.m.s. value of the difference of both fields is only 1.6 \%. The same coincidence is found for $\alpha=45^\circ,90^\circ$, $q_P=0.012$.\\
The temporal evolution of the magnetic and kinetic energies for both the stabilized and the destabilized background field configurations are shown in ~\figref{fig:mageq_stab} and ~\figref{fig:mageq_instab}:
\begin{figure*}[t]
\begin{tabular}{@{\hspace{0.5cm}}c@{\hspace{-.1cm}}c@{\hspace{3mm}}}
\multicolumn{2}{c}{\includegraphics[width=1.0\textwidth]{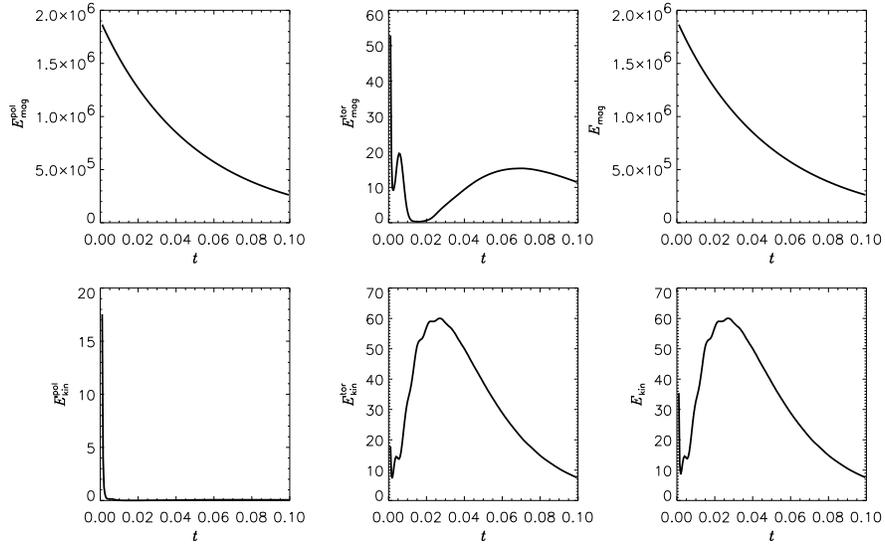}}  \\
\end{tabular} \caption{Temporal evolution of the disturbed dipolar magnetostatic equilibrium for $q_P$=0.12, $\Pm=1$, $\alpha=0$ (a stable case).  Time is in units of $\tau_\tOhm$, energy in units of $\rho\eta^2/R^2$. Subscripts ``kin'' and ``mag'' refer to velocity and magnetic--field related quantities, respectively.  The magnetic and kinetic energies, $E_{\mathrm{mag}}$ and $E_{\mathrm{kin}}$, are each further subdivided in their poloidal and toroidal parts.}
\label{fig:mageq_stab} 
\end{figure*}
\begin{figure*}
\begin{tabular}{@{\hspace{.5cm}}c@{\hspace{-.1cm}}c@{\hspace{3mm}}}
\multicolumn{2}{c}{\hspace*{-.2cm}\includegraphics[width=1.0\textwidth]{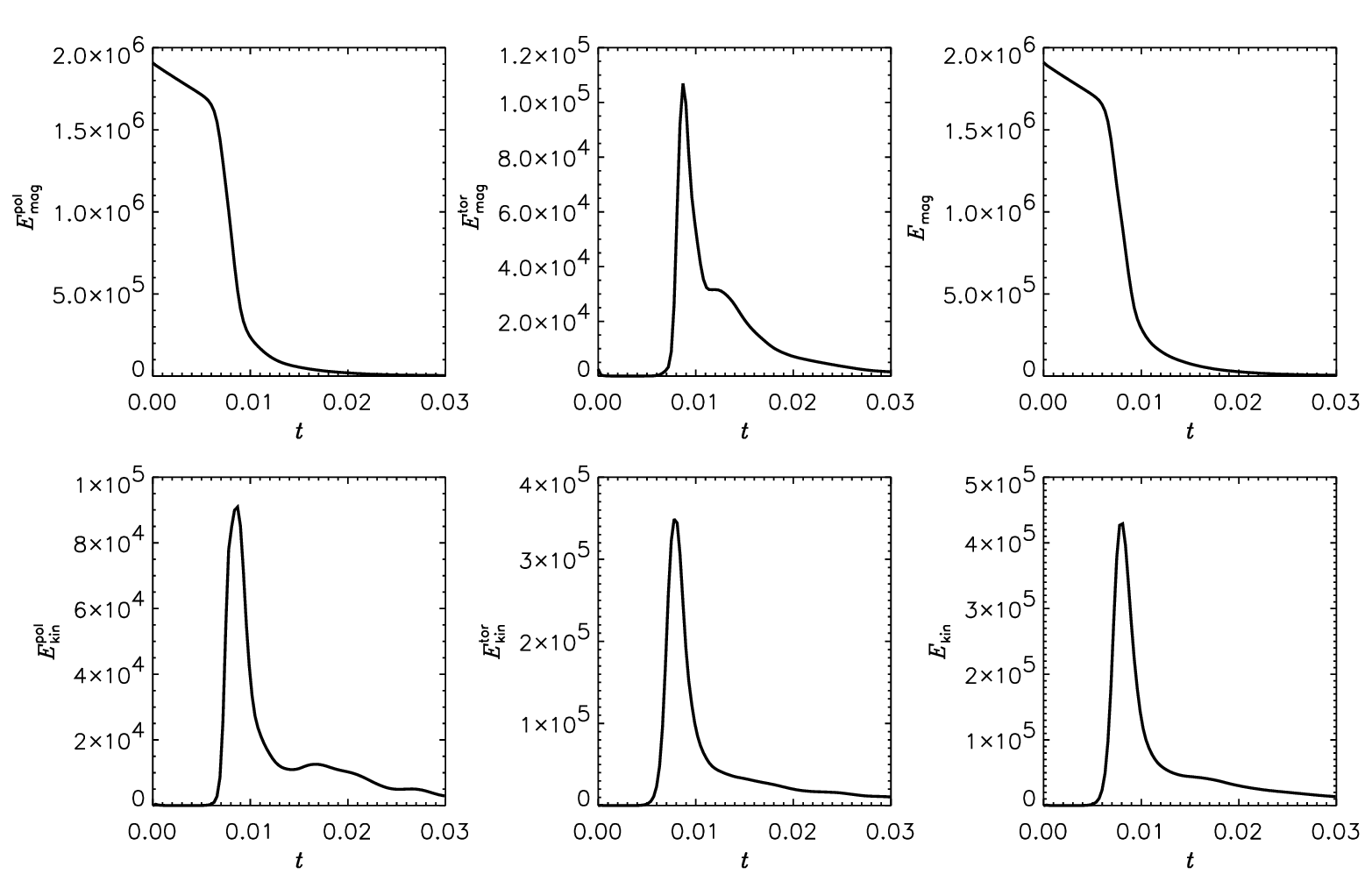}}\\
\end{tabular}
\caption{Temporal evolution of the disturbed dipolar magnetostatic equilibrium. $q_P$=12, $\Pm=1$, $\alpha=0$ (an unstable case). For further explanations see \figref{fig:mageq_stab}. Note, that using the parameters \eqref{denorm} $t=0.01$ corresponds to 0.5 seconds!}
\label{fig:mageq_instab}
\end{figure*}
A comparison of ~\figref{fig:mageq_stab} with ~\figref{fig:mageq_instab} shows that while for sufficiently fast rotation the magnetic energy is decreased to $15\%$ of its initial value (which corresponds to a decrease of the field strength from $1.71\cdot 10^{15}$ G to $7.1\cdot 10^{14}$ G) in the unstable , slower rotating NS, $99.99\%$ of the magnetic energy has been redistributed into kinetic one and is finally dissipated into heat, so that the remaining NS magnetic field has a strength of $1.71\cdot 10^{13}$ G, typical for the majority of young radio pulsars.\\
Concluding, the effect of the MHD instability occuring immediately after the birth of NSs with ultra--strong dipolar fields is that 
\begin{enumerate}
\item those whose rotation period is less than $\sim 6$ms and whose magnetic inclination angle $\alpha$ is smaller than $\sim 45^\circ$ will retain their extremely large surface field strength and appear, after a rapid spin down, as magnetars;
\item those which rotate less rapidly, say with $P \gtrsim 6$ ms and/or for which is $\alpha \gtrsim 45^\circ$ will lose almost all of their inborn magnetic energy and appear as radio pulsars.
\end{enumerate}
It turns out that rotation is likely to be the only stabilizing agent which allows of the existence of magnetars whereas the stable configurations found by Braithwaite \& Spruit (2006) are less suited to support such strong surface fields; due to their concentration to the very inner core region it demands too large field strengths for typical magnetar surface fields. Thus, the much smaller number of observed magnetars in comparison with that of all the other observed realizations of NSs may be a consequence of the fact that only a small fraction of all new--born NSs are rotating as fast as or faster than $P\sim 6$ ms.

\section{Fallback Accretion, Submergence and Rediffusion: Pulsar or Radioquiet Neutron Star?} \label{sec:2}

\begin{figure}[h]
\centerline{\psfig{file=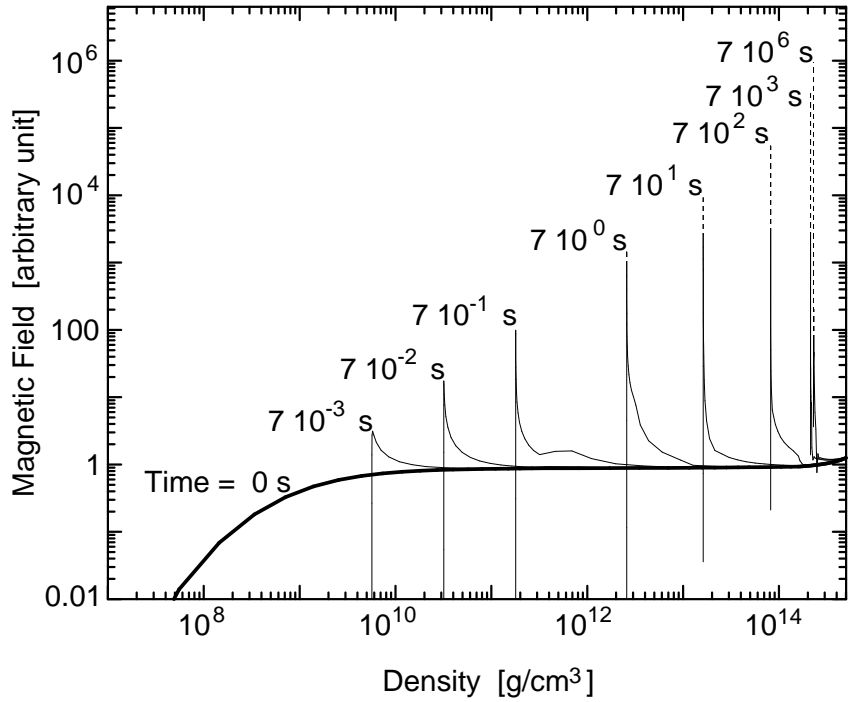,width=12cm,clip=} } 
\caption{Distribution of the angle averaged magnetic field strength in the NS as a function of time for the fallback rate estimated for SN1987A. Initial time corresponds to the beginning of the accretion phase. In less than two hours the initial field is submerged down to the crust--core interface. Notice that the maximum value attained by the field depends on its initial value at low density since the low density region is the most strongly sompressed. The calculation assumes ideal MHD, but these zones of highly compressed field have very small length scale and thus a very small ohmic diffusion time: these fields, shown as dashed lines, will eventually washed out by diffusion when time becomes comparable to the ohmic diffusion time.}
 \label{fig:submergence}
\end{figure}

\begin{figure}[h]
\centerline{\psfig{file=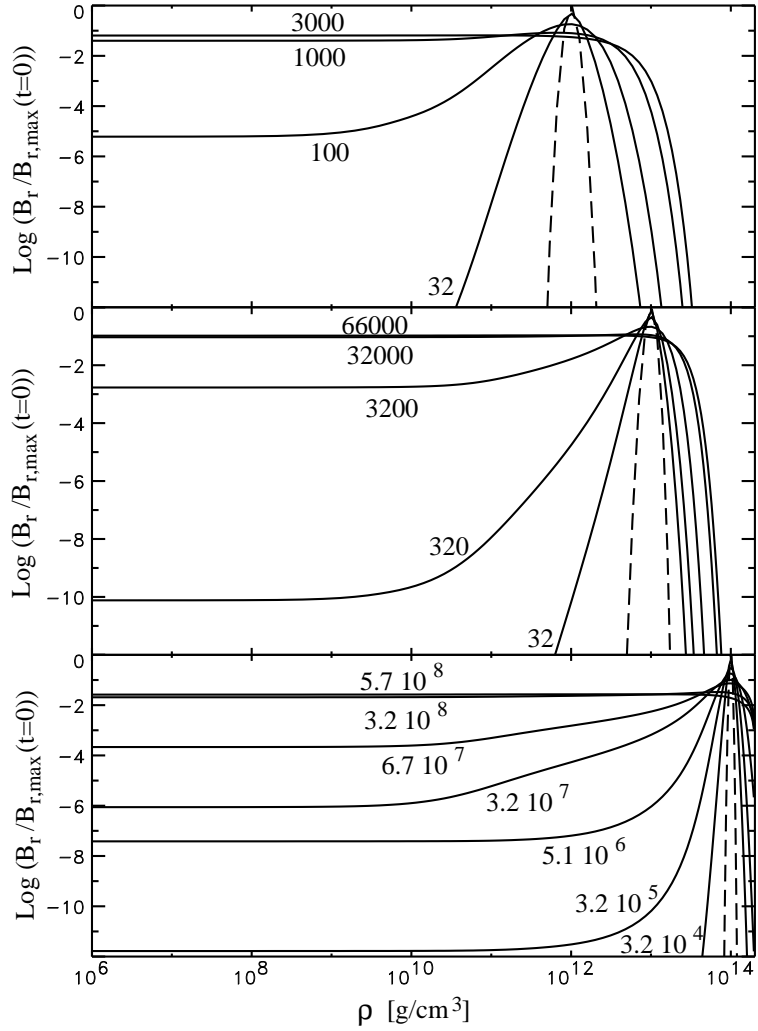,width=10cm,clip=} } 
\caption{Rediffusion of the interior magnetic field after the hypercritical accretion phase for three different submergence depths. Only the radial component is plotted. The assumed initial field location is shown by dashed lines. The ages of the star are indicated on the lines.}
\label{fig:rediffusion}
\end{figure}

Whether and if so how much fallback accretion can change the magnetic field evolution qualitatively depends on two questions:
\begin{itemize}
\item Can the fallback matter reach the NS surface at all, i.e. is the dynamical pressure of the matter flow stronger than the  pressure of NS's magnetic field which exists after the first $\sim 10$ seconds of NS's life (see~\sectref{sec:1}), and is the rotation sufficiently fast that the rotating dipole acts like a propeller?
\item If the fallback matter reaches the surface, how deep can the magnetic field be submerged and how fast can it rediffuse to the surface up to its strength before the 
hypercritical accretion started?
\end{itemize}
For other factors which may either reduce the rate of fallback accretion or prevent it at 
all (decay of $^{56}$Ni and $^{56}$Co, rapid rotation, kick velocity) see Geppert et al. (1999).
A rough estimate with respect to the first question is whether the \Alf{} radius, determined by 
the equilibrium of the pressure of the dipolar field and the ram pressure of the 
gravitationally captured fallback matter, is larger or smaller than the NS radius $R$. 
\begin{equation}
R_{A} = \left( \frac{R^{6} B^{2}}{4 \dot{M} \sqrt{GM}} \right)^{2/7}
\approx 1.3 \times 10^{5} \left( \frac{B_{12}^{2}}{\dot{M}_{350}}
\right)^{2/7} \; {\rm{cm}}, 
\label{R_A}
\end{equation}
where $B_{12} = B /10^{12}$G and $\dot{M}_{350} = \dot{M}/350
M_{\odot}$yr$^{-1}$. The accretion rate of $350 M_{\odot}$yr$^{-1}$ was estimated for the initial accretion on to the NS in SN 1987A (Chevalier (1989)); the factor in Eq.~\eqref{R_A} is given for a NS 
with $M=1.4M_{\odot}$ and $R=10^{6}$ cm. Clearly, the majority of NSs, having after the period of MHD instabilities "only" surface field strength in the order of $10^{12}$ G, will suffer from fallback if the accretion rate is as huge as in case of SN 1987A. If, however, the field strength after that periods is $\gtrsim 3\cdot 10^{13}$ G and the fallback accretion rate is only one tenth as strong as in the case of SN 1987A, the \Alf{} radius $R_{A} \approx 1.75\cdot 10^6$ cm, just the radius of a NS with quite stiff equation of state (EoS). In that case the submergence of the field can be attenuated drastically. Even for fallback as heavy as in case of SN 1987A, a magnetar field ($\gtrsim 10^{15}$ G) ensures $R_{A} \approx 6.9\cdot 10^6$ cm, a precondition for preventing fallback accretion.\\
Another condition to impede fallback concerns sufficient fast rotation. The rotation period which separates the propeller from the accretor regime is given by the so--called equilibrium period $P_{\mathrm{eq}}$ (Alpar 2001). Thus, the NS is in the propeller regime and can eject the inflowing fallback as long as
\begin{equation}
\left(\frac{2\pi}{P}\right)^2 R_{A}^3 > GM \approx 1.9\cdot 10^{25}\; {\rm{cm}^3} {\rm{s}^{-2}} \;
\label{prop_cond} ,
\end{equation}
i.e. for the magnetic field of standard pulsars $\sim 10^{12}$ G, even a rotation as fast as $P=10$ ms can not prevent accretion if it starts as heavy as in case of SN 1987 A (see Eq.~\eqref{R_A}). The same rotation combined with magnetar field strength, however, would drive the propeller mechanism. This mechanism prevents heavy accretion, but has an enourmously efficient braking effect (see e.g. Urpin et al. 1998). Since on the other hand the fallback accretion rate drops rapidly with time (in case of SN 1987A after the onset of accretion $\dot{M} \propto t^{-3/2}$ during the Bondi--accretion regime, after the transition to the dust--like regime $\dot{M} \propto t^{-5/3}$) it is possible that a magnetar field together with an rapid initial rotation may prevent the fallback accretion. For the majority of newborn NSs it is quite likely, that fallback accretion will appear, albeit not as heavy and field submerging as in case of SN 1987A. The onset of the powerful propeller regime which prevents the submergence of the magnetar field but spins the NS rapidly down 
might be one reason, why those NSs having magnetar field strength, the anomalous X--ray 
pulsar (AXPs) and the soft gamma repeaters (SGRs), are relatively young ($\sim 10^4$ yrs) and rotate so slow ($8 < P < 12$ s).\\
The question how deep the field can be submerged during fallback accretion and how fast it can rediffuse towards the surface has been addressed e.g. by Muslimov \& Page (1995), Geppert et al. (1999), and Cumming et al. (2001).\\
The magnetic field present at the surface of the NS when accretion stops will be the field which was present in the accreted matter and compressed. Following the standard hypothesis that the pulsar magnetic field is a fossile of the progenitor's core field, the accreted matter, being material of the progenitor's core too, could bring in a field comparable to the field already present in the NS, i.e., the NS may be born with a strong surface field.\\
However, the hypothesis that the fall-back matter brings in a well--ordered large scale field is questionable since there is still the possibility that this accreting matter has suffered a turbulent episode during which the plasma behaved as a diamagnet (Va\u{i}nshte\u{i}n \& Zel'dovich 1972) and its field could have been severely reduced, which would mean that the final surface field of the NS were also weak. In contradistinction, within the proto-NS dynamo scenario for the origin of NS magnetic fields (Thompson \& Duncan 1993) the core of the progenitor is only required to have a small field which will act as a seed for the dynamo action. In this case the field present at the NS surface after accretion will be small. The strength of the surface  magnetic field of a new-born NS which has undergone hypercritical accretion may thus be very different if its magnetic
field is fossile or of proto-NS dynamo origin. Hence, the assumption that the accreting matter is only weakly, magnetized, is natural within the proto-NS dynamo scenario and may also be compatible with the fossile field hypothesis.\\
How deep the field will be submerged depends on the details of the supernova explosion as well as on the magnetic field strength and on the rotation rate of the NS when the fallback hits the surface. The submergence process in a NS whose field is that of "standard" pulsars and for the fallback parameters of SN 1987A  (for details see Geppert et al. 1999) is shown in \figref{fig:submergence}.\\
When the accretion has ceased, the field starts re--diffusion back toward the surface.
This problem has been considered by Muslimov \& Page (1995) in the case of very shallow submergence, i.e. the total amount of accreted matter $M_{\mathrm{acc}} \sim 10^{-5} M_{\odot}$. They showed that after a few hundred years the surface field strength becomes comparable to the interior one, resulting in a delayed switch--on of the pulsar.
For a typical type II SN Chevalier (1989) estimated that the
accreted mass should be at least $100$ times smaller than in SN 1987A, 
i.e. less than $10^{-3} M_{\odot}$.\\
The rediffusion process in isolated NSs is solely determined by the conductive properties of the crustal matter. This, in turn, depends on the cooling scenario and on the impurity content $Q$ of the crust. The rediffusion processes shown in \figref{fig:rediffusion} are based on the standard (slow) cooling scenario (see Page et al. 2005) and an impurity content of $Q=0.01$. The latter is
controversially discussed, see Jones (2004) who argues in favor of a much larger $Q$ which would accelerate the rediffusion.
The final depths into which the field has been submerged after accretion of a certain amount of fallback matter (see Fig. 2 in Geppert et al. 1999) correspond roughly to a rediffusion time of about
$10^3$ ($3 \times 10^4$) yrs for $M_{\mathrm{acc}} \sim 10^{-4}$ ($\sim 10^{-3}$) $M_{\odot}$, while for  $M_{\mathrm{acc}} \sim 10^{-2} M_{\odot}$ rediffusion takes more than $10^8$ yrs. Moreover, in the case of $M_{\mathrm{acc}} \sim 10^{-1} M_{\odot}$ the rediffusion time will exceed the Hubble time and, as far as SN1987A is concerned, it is likely that a pulsar will never be seen in it.\\
This re--magnetization scenario relies upon the assumption that the accreted matter is weakly magnetized, either because the progenitor's core had a very weak magnetic field or because the explosion and/or accretion process demagnetized it.\\
There remains also the possibility that immediately after the fallback a mechanism generates a strong field in the very surface layers based on a thermoelectric instability (\sectref{sec:3}) driven by the strong temperature gradient in the outer crust. Even stronger temperature gradients may appear transiently during the thermal relaxation of the young NS in deeper crustal layers (Gnedin et al. 2001), which would be preferred locations for the transfer of thermal in to magnetic energy. In that case, the pulsar in the remnant of SN 1987A could be switched on relatively soon.\\

\section{Thermoelectric Instabilities: Strong Fields Despite Deep Submergence?} \label{sec:3}

Wherever in nature large temperature gradients are maintained in a medium of sufficiently high conductivity, a suitably structured arbitrarily weak magnetic seed field can be amplified by an instability which is based on two thermomagnetic effects, the thermo-Hall effect, by which the magnetic field affects the heat flux (see \sectref{sec:4.2}) and the thermoelectric effect, by which a temperature gradient creates an e.m.f. (battery effect). This instability is e.g. used to confine plasmas in thermonuclear reactor devices (Winterberg 2005).\\
The existence of huge temperature gradients is one of the many superlatives which are assigned to NSs. For the first time Dolginov \& Urpin (1980) studied the possibility of an thermomagnetic instability in the cores of white dwarfs. Soon it became clear that in the envelopes of NSs much larger temperature gradients are prevalent (Gudmundsson et al. 1983) which, together with the high
electric conductivity may guarantee that the field generation overwhelms the ohmic diffusion. Blandford et al. (1983) considered thermoelectric field amplification  in the solid crust which should via Lorentz forces drive a dynamo process in the liquid layer above the solid. Urpin et al. (1986)
showed that the thermomagnetic instability may act efficiently in the liquid layer only and that a sufficiently fast rotation is necessary to keep the instability alive. The latter condition is fulfilled by the vast majority of young NSs.\\
The basic scenario is sketched in ~\figref{fig:TEI0} and can be described as follows:
\begin{figure}
\centerline{\psfig{file=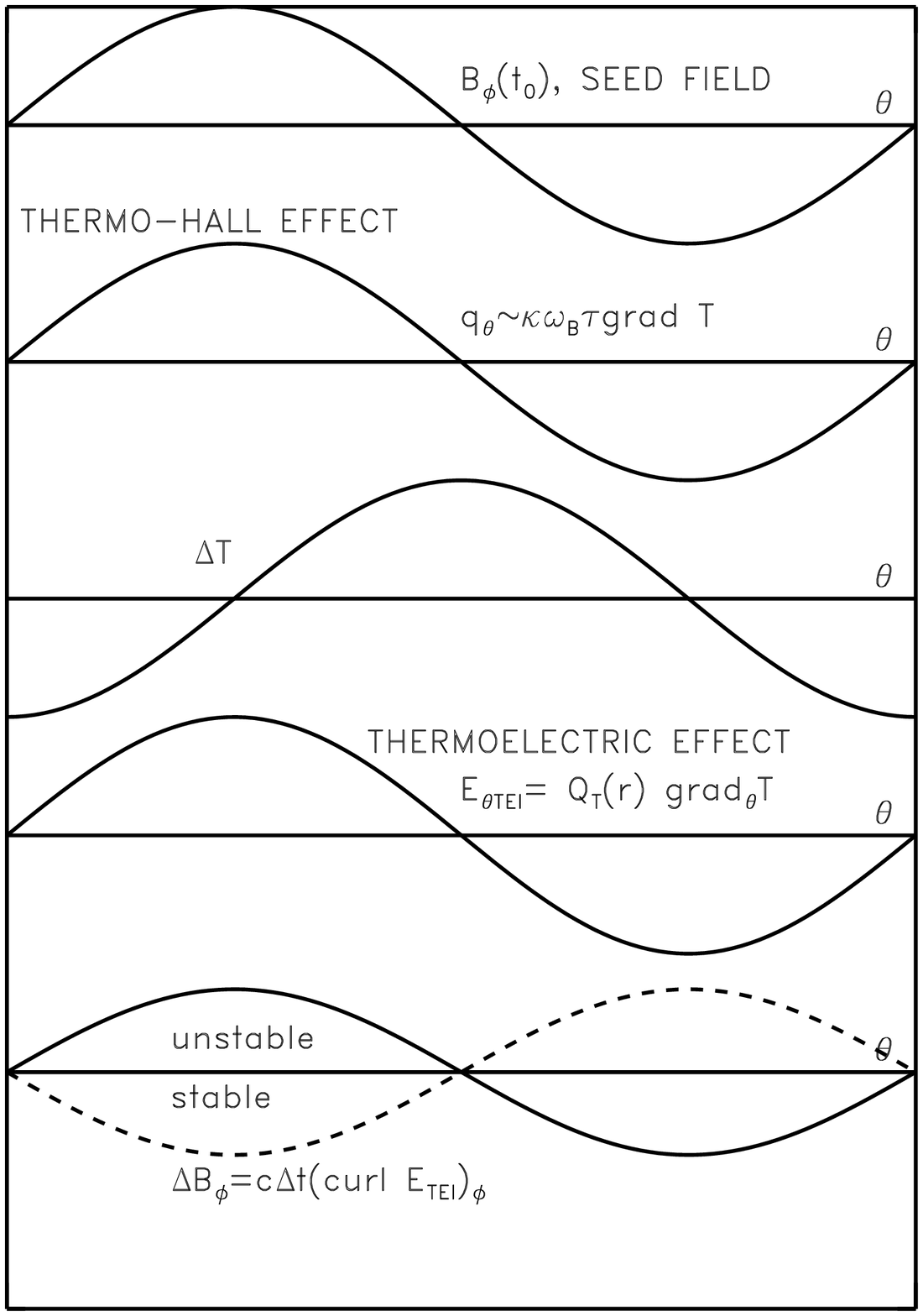,width=6cm,clip=} } 
\caption{Schematic sketch of the mechanism of the thermoelectric instability. During the linear stage of the instability only toroidal seed fields can be amplified (according Fig.1 of Urpin et al. 1986). }
\label{fig:TEI0}
\end{figure} 
A pre--existing  small scaled toroidal (in axisymmetry: azimuthal) component of the NS magnetic field in the liquid creates via the thermo--Hall effect by means of the strong radial temperature gradient thermal flux variations in meridional direction having approximately the same scale length as the seed field. This meridional heat flux causes temperature variations and the thermoelectric effect
generates by them an additional electric field also pointing in meridional direction. Due to the
non--uniformity of the liquid layer in radial direction that electric field has a curl component in azimuthal direction which, under suitable  conditions, may amplify the seed field.\\
The set of equations which govern the thermoelectric instability is:

\begin{equation}
 \begin{alignedat}{2}
\parder{\Bvec}{t} &= -\curl(\frac{c^2}{4\pi\sigma} \curl{\Bvec}) + c\;\grad{Q_T}\times \grad{T} + \curl\left[(\vec{v_j} + \vec{v_{TD}})\times \Bvec \right] \; ,\\
0 &= \dv \left[(\kappa_{\mathrm{ei}} + \kappa_{\mathrm{rad}})\grad{T} + \kappa_{\mathrm{ei}} \omega_B \tau (\vec{b}\times \grad{T})\right] \; .\\
 \end{alignedat}
\label{TEI}
\end{equation}
The first term in the induction equation describes the ohmic diffusion of the field, the second term is the battery term. Its strength is determined by the temperature gradient and the gradient of the thermopower $Q_T$. Because all transport coefficients are in good approximation dependent on the radial coordinate only as long as $\omega_B \tau < 1$, the battery term can amplify a seed field only if the temperature gradient has, besides its strong radial component, a meridional one too. The third term resembles the usual advection term. Here, however, the velocity is not the hydrodynamical motion, perhaps affected by the Lorentz force. Since the instability starts with weak seed fields, the coupling to the hydrodynamics is neglected. This assumption becomes wrong only if the field strength exceeds $\sim 10^{12}$ G. Instead, the velocity consists here of the thermal drift $\vec{v_{TD}}$ which describes the drift of the magnetic field in the liquid caused by the temperature gradient and is a consequence of the thermoelectric effect. The electron mobility ($\vec{v_j} \propto \curl{\Bvec}$) is responsible for the Hall--drift; it makes the induction equation nonlinear in $\Bvec$ while the thermal drift together with the battery effect as well as the thermo--Hall effect couples the field evolution to the thermal one.\\
The thermal conductivities $\kappa_{\mathrm{ei}}$ and $\kappa_{\mathrm{rad}}$ correspond to the heat transport due to electron--ion collisions and to radiation, respectively, where the latter dominates with decreasing density in the liquid layer. For the field strength expected to appear at this stage of field evolution the radiative conductivity will not be affected by the field. It can only influence the electron--ion collisions which, together with the magnetization parameter and the temperature gradient determine the relative importance of the thermo--Hall effect, coupling the heat flux to the field ($\vec{b}$ is the unit vector of $\Bvec$). For details of the derivation of Eqs.~\eqref{TEI} see Geppert \& Wiebicke (1991).\\
Note, if there is initially a purely radial temperature gradient, initially only the toroidal component of the seed field can be amplified. Any amplification of the poloidal field component, which forms the dipole field outside the NS, is only possible via nonlinear interactions of the poloidal and toroidal field components and each of them with the temperature variations.\\
In a series of studies Geppert \& Wiebicke (see Wiebicke \& Geppert 1996 and references therein) tried to follow the evolution from a weak toroidal seed field to a poloidal field of observed pulsar strength, but they failed. They could show the scheme of thermoelectric field generation in the surface layers of young NSs which is characterized by a rapid  growth of small scale toroidal field components in less than $10$ yrs saturating at field strengths $\sim 10^{13}$ G, provided the surface temperature is $\gtrsim 3\cdot 10^{6}$ K. Below that surface temperature the temperature gradient in the liquid crust becomes too flat and ohmic decay and/or the Hall drift will dominate the field evolution. During the exponential growth of the small--scale toroidal field modes nonlinear (quadratic) interactions drive an (twice as) fast growth of large scale (say quadrupolar) toroidal fields which reach in about $100\ldots 1000$ yrs field strength of $10^{11\ldots 12}$ G (see \figref{fig:TEI1}). While after about $10$ years the exponential growth of the small--scale modes saturate, the large--scale modes are still growing.\\
\begin{figure}
\centerline{\psfig{file=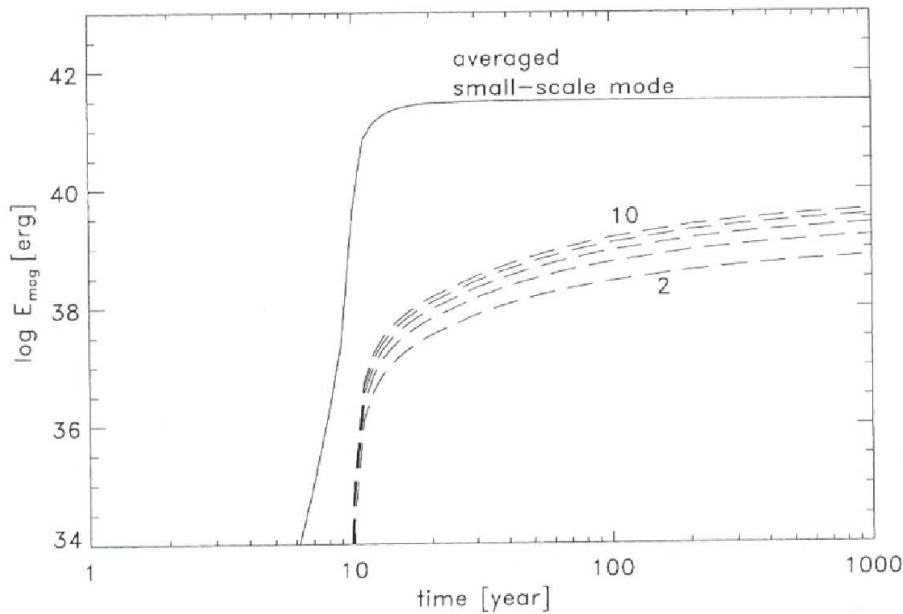,width=12cm,clip=} } 
\caption{Magnetic energy content of the toroidal field modes as function of time. The result has been obtained for a NS model with $T_{\mathrm{s}}= 5\cdot 10^6$ K by a nonlinear calculation which couples the 5 largest scale modes (dashed lines, even multipolarities n= 2,4,6,8,10) to each other and the small scale "locomotives". The energy content of the large scale modes remains 2 orders of magnitude smaller than that of the "collective" fastest growing small scale mode. For details see Geppert \& Wiebicke (1995).}
\label{fig:TEI1}
\end{figure} 
\noindent However, the growth of the large--scale toroidal modes after saturation of the small--scale modes as shown in ~\figref{fig:TEI1} is questionable. There are two reasons why the modelling of their growth and that of the poloidal field component, fed by the rapidly growing small--scale toroidal components, did not return correct results. Firstly, the decoupling of the field from the hydrodynamic motions is not justified when the field attains strengths  $\gtrsim 10^{12}$ G. Then, Lorentz forces may drive matter circulations which can act dynamo--like and amplify the poloidal component of the seed field, too. Secondly, it is well possible that the Hall--instability sets in as discussed in \sectref{sec:4.1}. An indication of this scenario is that as soon as the toroidal field component exceeded $\approx 10^{12}$ G, the Hall--drift  caused a rapid growth of smaller scaled components and the code crashed.\\
Although the complete thermoelectric field generation process in the crust is by far not yet understood and both the adding of the equations of hydrodynamics and the numerical handling of the Hall instability are quite challenging complex problems, I would like to mention a place and a situation in the NS, where the thermoelectric instability my act even more efficiently than in the outer liquid layer of the crust.\\
Gnedin et al. (2001) studied the thermal relaxation in young NSs which proceeds when during the first $100$ yrs the core and the outer crust of the NS cools by neutrino emission faster than the bulk of the crust in the range $5\cdot 10^{11} < \rho < 2\cdot 10^{14}$ g cm$^{-3}$ for a standard (slow) cooling scenario. This causes naturally two temperature gradients just around these limiting densities (see \figref{fig:TEI2}). 
\begin{figure}
\centerline{\psfig{file=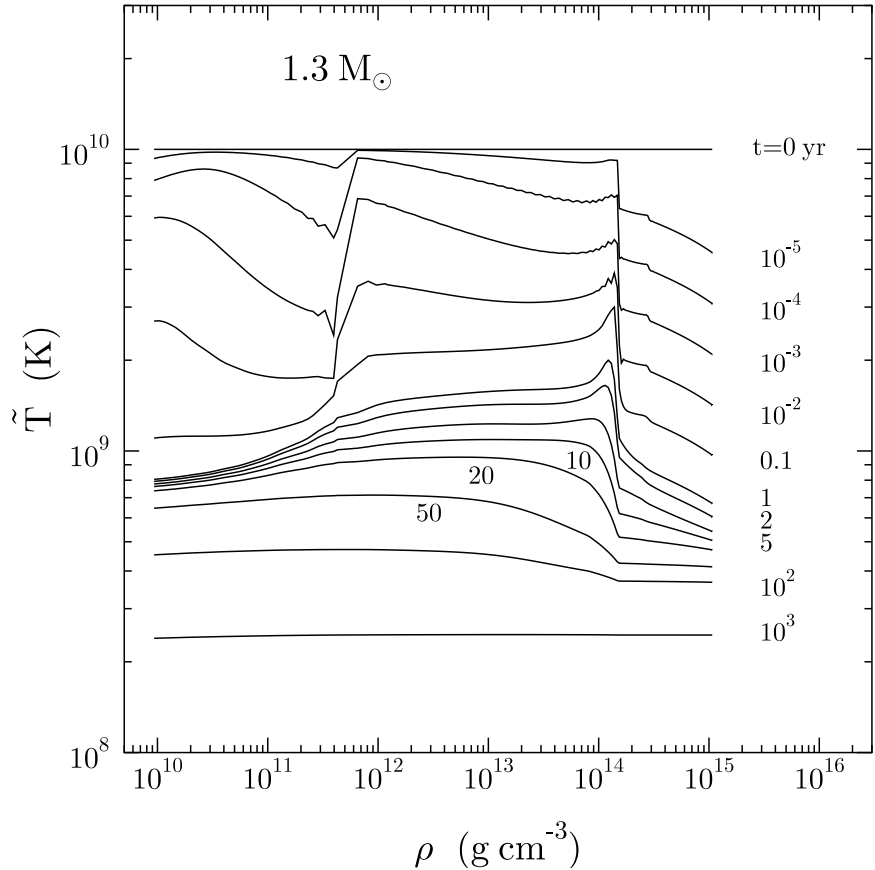,width=12cm,clip=} } 
\caption{Temperature profile in a $1.3 M_{\odot}$ NS without superfluidity effects as shown by Gnedin et al. (2001) depending on the age which is indicated by the numbers next to the curves.}
\label{fig:TEI2}
\end{figure} 
The temperature gradient, e.g. at the crust--core boundary is in the order of $2.5\cdot 10^6$ K cm$^{-1}$, i.e. about $50\ldots 100$ times stronger than in the outer liquid layer discussed above. Since the growth time scales are proportional to the square of the inverse of the temperature gradient (Dolginov\& Urpin 1980) and the growth time in the outer crust is for surface temperatures $T_{\mathrm{s}} \gtrsim 3\cdot 10^6$ K in the order of $50\ldots 100$ days, one can expect that, for the instability acting just above the core, the growth time is $\sim 500\ldots 2000$ seconds only. Depending on the cooling, which determines the onset of crystallization, there could be conditions realized, under which a large magnetic field is very fast generated. Another support for the instability is the much larger electric conductivity in comparison with that appearing in the outer crust. Additionally the inverse direction of the temperature gradient may help, which tends to drive the field in regions with even larger conductivity. Perhaps, the rapidly growing magnetic field in the inner crust will prevent the liquid matter for a while from crystallization. Moreover, strong toroidal fields present in the vicinity of the crust--core boundary are necessary to explain the existence of small hot polar regions as discussed in \sectref{sec:4.2}. I believe it worthwile to consider the possibility, that very early on, just after the MHD instabilities in the newborn NS ceased, deep in the star strong fields may be created by use of extreme temperature gradients. It seems, however, that for this purpose the hydrodynamics has to be coupled to Eqs.~\eqref{TEI}.\\

\section{Large Magnetization Parameters: Hall--Drift Induced Instabilities and Strongly Anisotropic Surface Temperatures?} \label{sec:4}

As already mentioned in \sectref{sec:0}, a magnetization parameter $\omega_B \tau$ exceeding unity will change the magnetic field evolution  and cooling history, causing various, possibly observable consequences. The reason is that both the electric and the heat conductivity are in the presence of a magnetic field no longer only {\it scalar} functions of density, temperature, chemical composition and other quantities, but become {\it tensors}. If $\omega_B \tau \gg 1$ the tensor components perpendicular to the magnetic field lines will be suppressed dramatically. Therefore, the induction equation becomes nonlinear  and the heat transport will proceed almost exclusively parallel to the magnetic field. While the nonlinear induction equation comprises the multifaceted effects of the Hall drift, a significant deviation from isotropic heat transfer through the crust affects the surface temperature distribution of thermally emitting isolated NSs.

\subsection{Hall--Drift in the Crust}\label{sec:4.1}
\begin{figure}
\begin{center}
\epsfig{file=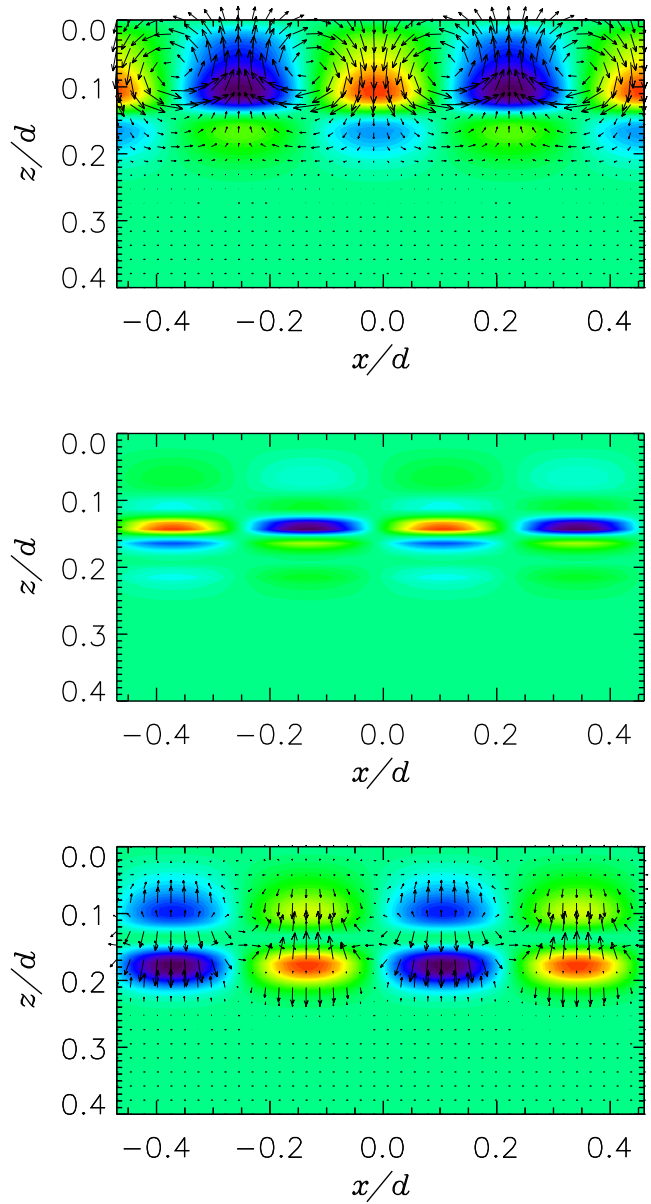,width=11cm}   
\end{center}
\caption{Consequences of the Hall drift. Upper panel: typical structure of the small scale field generated on a growth time of $3\cdot 10^3$ years by the Hall instability from a large scale toroidal crustal field of $3\cdot 10^{13}$ G. The thickness of the crust is $d \approx 3800$ m. Thus, the typical meridional and azimuthal scale of the perurbations is about 1 km. The field is concentrated in a depth of about $400$ m below the NS surface $z=0$. Colour encoding corresponds to the azimuthal field component (for details see Rheinhardt et al. 2004). Middle panel: Perturbation of the Joule heat sources density $\propto 2\curl{\Bvec_0}\times \curl{\delta\bvec}$ in arbitrary units corresponding to the perturbation field shown in the upper panel. Green to red - positive, green to blue - negative deviations from the background heat sources. Lower panel: Lorentz force density perturbations $\propto \curl{\Bvec_0}\times \delta\bvec + \curl{\delta\bvec}\times \Bvec_0$ corresponding to the perturbation field shown in the upper panel. The arrows denote radial and meridional force components, colour encoded is the azimuthal force component.}
\label{fig:Hall_Feld}
\end{figure}

Simultaneously with the tensorial character of the electric conductivity, 
two nonlinear effects are introduced into Ohm's law: the Hall drift and the ambipolar diffusion. However, if the conducting matter consists of electrons and only one sort of 
ions, but no neutral particles take part in the transport processes
the ambipolar diffusion is absent (cf. Yakovlev \& Shalybkov 1991). Such a
situation is realized in crystallized crusts of NSs
and/or in their cores if the neutrons are superfluid, but the protons are normal and the electrons may therefore collide with protons but effectively not with the neutrons.\\
Many authors discuss the consequences of the Hall drift in isolated NSs, see e.g. Haensel et al. (1990), Goldreich \& Reisenegger (1992), Naito \& Kojima (1994), Urpin \& Shalybkov (1995), Shalybkov \& Urpin (1997),   Va\u{i}nshte\u{i}n et al. (2000), Hollerbach \& R{\"u}diger (2004) and Cumming et al. (2004) and references therein. They discussed the redistribution of magnetic energy from an initially large--scaled (e.g. dipolar) field into small--scale components due to the nonlinear Hall term. Though the Hall drift itself is a non--dissipative process, the tendency to redistribute the magnetic energy into small scales may accelerate the field decay considerably.\\
Va\u{i}nshte\u{i}n et al. (2000) found that the Hall drift creates current sheets in configurations where a large density gradient exists. These current sheets can be sites for rapid ohmic dissipation of magnetic energy. Since crusts of NSs have a very large density gradient ($\sim 10^{14}$g cm$^{-3}/ 10^5$ cm) Va\u{i}nshte\u{i}n et al. (2000) concluded that in current sheets created by a crustal magnetic field this could decay an timescales of $3000\ldots 30$ years, depending on the location of the current sheets within the crust.\\
When starting with a large scale magnetic field the {\it Hall cascade} derived by Goldreich \& Reisenegger (1992) will generate small scale field
components down to a scalelength $l_{crit}$, where the ohmic dissipation begins
to dominate the Hall drift. This cascade, however, can be accompanied or superimposed by a non--local (in the spectral space) magnetic enery transfer from a slowly, (ohmically) decaying, larger scale background field into smaller scale components. This {\it Hall instability} may have observable consequences (Rheinhardt\& Geppert 2002 and Geppert \& Rheinhardt 2002). Moreover, the Hall instability and/or cascade are well conceivable processes which produce the strong surface field components of  smaller scale ($l \sim R/10$), necessary for the pulsar mechanism to work (Geppert et al. 2003).\\
The occurrence of the Hall instability is based on certain properties of the electric currents maintaining the background field: the motion of the elctrons which create the currents must show a sufficiently strong shear (Cumming et al. (2004), Rheinhardt \& Geppert (2005)). A linear stability analysis performed in a plan--parallel slab assuming for simplicity constancy  for the transport coefficients reveals the mechanism of the instability.\\
With $\Bvec=\Bvec_0+\delta\bvec$, where $\Bvec_0$ denotes the background field (chosen such, that its Lorentz force is a gradient) and $\delta\bvec$ a small perturbation, the linearized dimensionless induction equation
\begin{equation}
\dot{\delta\bvec} = \Delta\delta\bvec - \curl (\,\curl\Bvec_0
\times\delta\bvec\ + \curl\delta\bvec\times\Bvec_0\,)\;,\;\;\dv \delta\bvec = 0 
\label{indeqdimlesslin}
\end{equation}
describes the behaviour of the perturbations  of the reference state (for details see Rheinhardt \& Geppert 2002).
Along with the term $\curl\delta\bvec\times\Bvec_0$ which is
energy--conserving like the original Hall term $\curl\Bvec\times \Bvec$ here a second Hall term $\curl\Bvec_0\times\delta\bvec$ occurs
which may well deliver or consume energy (to/from $\delta\bvec$ !) since in general the integral $\int_V (\curl\Bvec_0\times\delta\bvec)\cdot \curl\delta\bvec\, dV$ will not vanish. This reflects the fact that Eq.~\eqref{indeqdimlesslin} describes 
the behavior of only a part of the total magnetic field. Actually, perturbations
may grow only on expense of the energy stored in the background field.\\
Performing a standard stability analysis, the perturbations $\delta\bvec \propto \exp{pt}$ are found to have for a certain range of background field strengths positive growth rates $p$ which correspond to characteristic growth times of $10^3\ldots 10^5$ years; has the background field magnetar strength the growth time reduces to $\sim 10$ yrs. Note, that from Eq.~\eqref{indeqdimlesslin} the critical scale length below which ohmic dissiation dominates the Hall drift is $l_{crit} \le L/(\omega_B\tau)$ ($L$ being the scale length of the background field); at the same $l_{crit}$ the Hall cascade ceases.\\
This rapid transfer of magnetic energy may cause observable consequences. The drain of energy from the large scale background field, which determines the rotational evolution by magnetodipole radiation and stellar wind, weakens - at least episodically - the ability of that large scale field to spin down the NS. This should be reflected observationally by braking indices $n = 2 -P\ddot{P}/\dot{P}^2$ exceeding markedly the value $n=3$ for a constant dipole. Such values have been found for a number of radiopulsars as old as $10^5 \ldots 10^6$ years (Johnston \& Galloway 1999). Geppert \& Rheinhardt (2002) have shown that the Hall instability may reduce the dipolar field with a rate of $\sim 10^8$ G yr$^{-1}$, in coincidence with some of the observations.\\
Another consequence of the Hall cascade and/or instability is the generation of small scale field structures close to the NS surface, which automatically cause small scale Lorentz forces and Joule heating sources. A typical structure which may arise due to the Hall instability is shown in ~\figref{fig:Hall_Feld}. It is obtained by solving the Hall induction equation at a certain moment of the NS's cooling, reasonably assuming that for NSs older than $10^5$ yrs its cooling time scale is larger than the growth time of the Hall instability in case of $B_0 \gtrsim 10^{13}$ G. Moreover, in calculating that structure a realistic crustal density profiles has been applied. Thus, in comparison with Eq.~\eqref{indeqdimlesslin}, the Hall induction equation is not longer dimensionless and has an additional term.
\noindent For a background field of $3\cdot 10^{13}$ G the maximum growth time of the perturbations is in the order of $3\cdot 10^3$ years; it scales inversely with the backround field strength. The generated small scale poloidal field structures have just length scales as required by the Ruderman \& Sutherland pulsar model to drive the pulsar`s radio emission (Ruderman \& Sutherland (1975)). Since the currents which maintain these small scale fields are circulating in relatively low density crustal layers, they decay on a timescale of $\sim 10^6$ years after the Hall instability lost its power because the energy loss of the background field became so large that it is no longer unstable.\\
It is conceivable that these consequences of the Hall--instability  in case of magnetar background field strengths ($\ge 10^{15}$ G) are responsible both for the bursts observed in the SGRs and the thermal emission of SGRs and AXPs.\\

\subsection{Temperature Distribution in the Magnetized Crust}\label{sec:4.2}
\begin{figure}
\centerline{\psfig{file=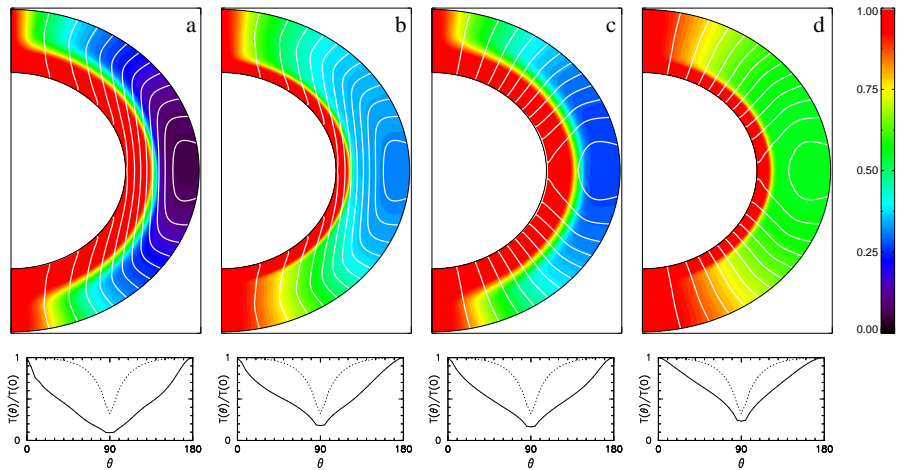,width=12cm,clip=} } 
\caption{Thermal structure of the NS crust between the crust--core interface and the bottom of the envelope at $\rho = 10^{10}$ g cm$^{-3}$. The radial scale of the crust is stretched by a factor of 5 for clarity. The magnetic field includes the three costituents $\vec{B}^\mathrm{crust}$, $\vec{B}^\mathrm{core}$ and $\vec{B}^\mathrm{tor}_0$. In panels a and b the crustal poloidal field dominates the core field ($B^{\mathrm{crust}}_0 = 7.5\cdot 10^{12}$G, $B^{\mathrm{core}}_0 = 2.5\cdot 10^{12}$G) while in panels c and d the core field is dominant ($B^{\mathrm{crust}}_0 = 2.5\cdot 10^{12}$G, $B^{\mathrm{core}}_0 = 7.5\cdot 10^{12}$G. In all panels $\vec{B}^\mathrm{tor}_0= 3\cdot 10^{15}$ G. Here, the index "0" denotes the polar surface (for the crustal and core poloidal field) and the maximum (for the crustal toroidal field) values. In the lower panels the full lines show the resulting surface temperature profiles $T_{\mathrm{s}}(\theta)$ and the dotted lines illustrate the same profile when an isothermal crust is assumed, i.e. if the magnetic field would influence the heat transfer in the envelope only.}
\label{fig:T_crust}
\end{figure}

\begin{figure}
\centerline{\psfig{file=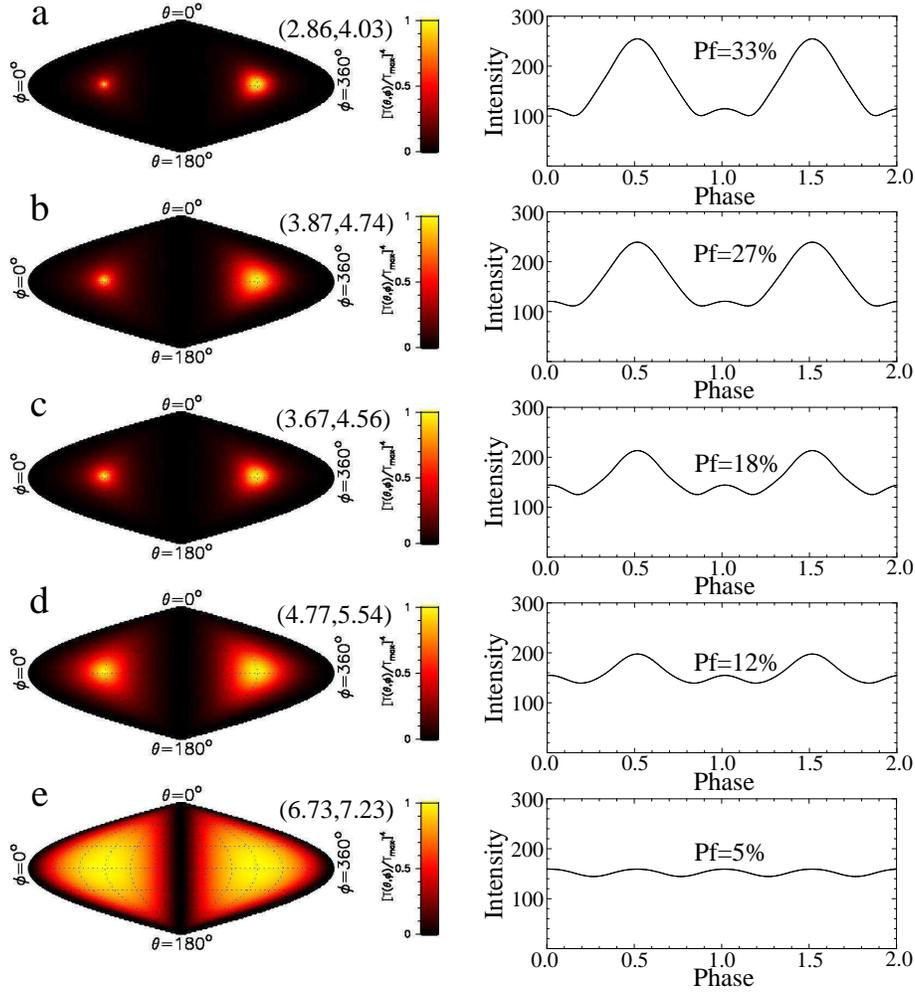,width=12cm,clip=} } 
\caption{Surface temperature distributions (left panels) in an area preserving--representation with a color scale following the emited flux ($\propto T^4$). Panels a to d use the internal field structures of the corresponding panels in Fig.~\protect\ref{fig:T_crust} while panel e assumes an isothermal crust. In contrast to Fig.~\protect\ref{fig:T_crust} the dipolar symmetry axis is in all cases oriented in the equatorial plane defined with respect to the rotation axis $\theta = 0,\pi$. The right panels show the resulting pulse profiles (in arbitrary units) which an observer, also located in the equatorial plane, would detect. In all cases the core temperature is the same but the star's distance has been adjusted to give the same average flux (see Fig.~\protect\ref{fig:Spectra}). Number pairs within parentheses give ($T_\mathrm{ave}$, $T_\mathrm{eff}$) with $T_\mathrm{ave}$ the optical flux and  $T_\mathrm{eff}$ for the X--ray flux (see Geppert et al.2006), resp., in units of $10^5$ K. All five models have almost the same maximum surface temperature $T_\mathrm{max} \simeq 8.45 \times 10^5$ K but different minimal temperatures.}
\label{fig:Tdistr_Litc}
\end{figure}

\begin{figure}
\centerline{\psfig{file=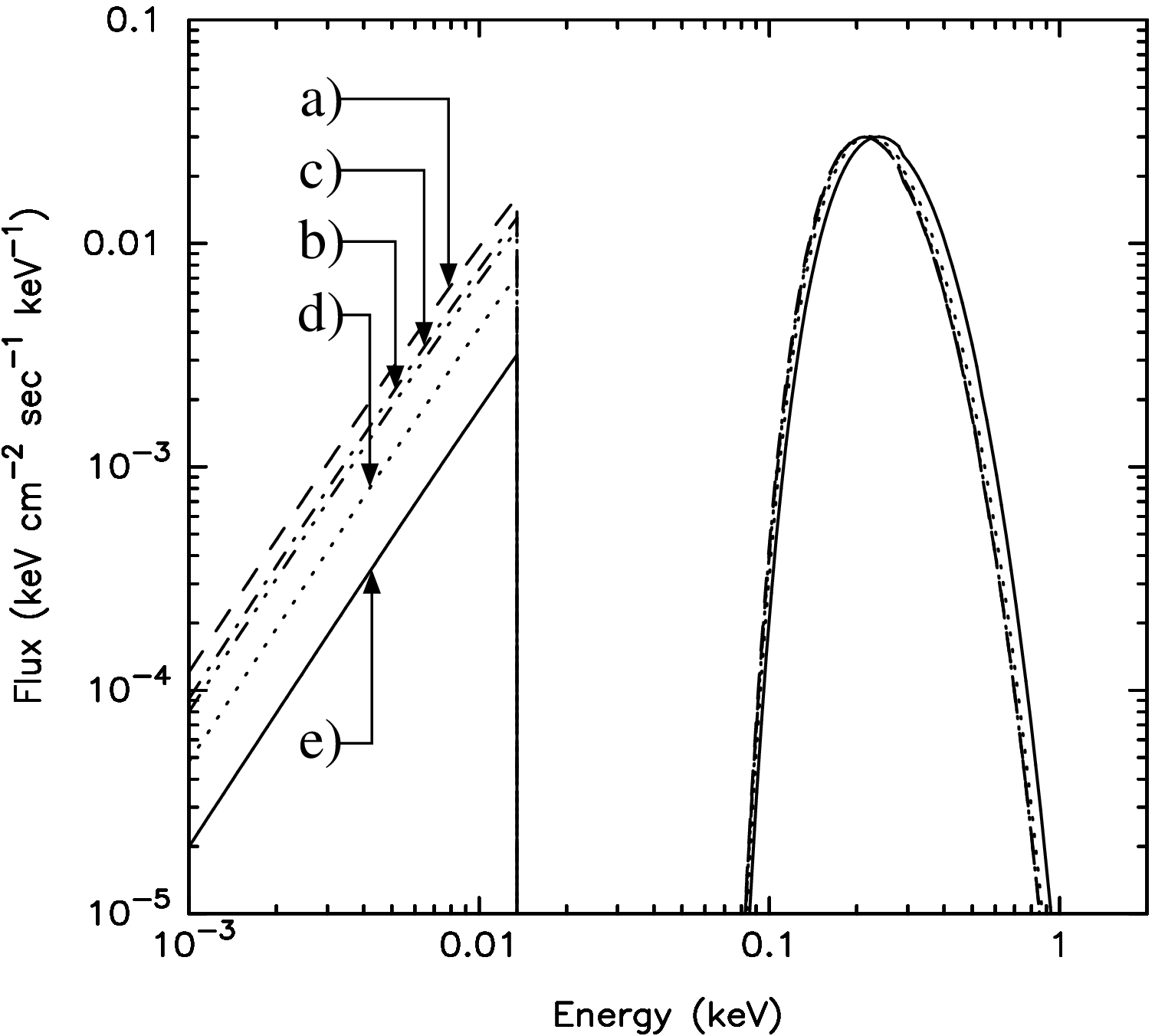,width=10cm,clip=} } 
\caption{Observable spectra for the five surface temperature distributions and pulse profiles, ``a'' to ``e'', shown in Fig.~\protect\ref{fig:Tdistr_Litc}. 
The stars, with radius $R=11.4$ km (radius seen at infinity $R_\infty = 14.28$ km) and $M=1.4 M_\odot$, are assumed to be at distances of 100, 142, 131, 202, and 220 pc, resp., to produce almost identical observable spectra in the X-ray band 
(column density $N_{\mathrm{H}} = 1 \times 10^{20}$ cm$^{-2}$ for interstellar absorption) 
but resulting in significantly different fluxes in the optical range.}
\label{fig:Spectra}
\end{figure}

In the crust of NSs electrons are the by far dominating carriers of the heat flow. By collisions with impurities and phonons in the crystallized crust and with ions in its liquid layer they transfer the heat following the temperature gradient from the core through the crust and its envelope towards the surface where it is finally irradiated. In case of weak magnetization, the heat will be transferred almost isotropically and a uniform surface temperature $T_{\mathrm{s}}$ would be seen by an observer.\\
Observations of many isolated NSs, however, indicate with great significance that the surface  temperature $T_{\mathrm{s}}$ is not uniform but has (in some cases even large) meridional gradients. For slowly rotating, radio quiet, isolated NSs as the ``Magnificent Seven'' (see review of Haberl in this volume) the magnetic field can be the source of a significant deviation from an isotropic surface temperature distribution. Outstanding common features as seen for the ``Magnificent Seven'' are the apparent smallness of their radii derived from X--ray spectra, the slow rotation, and the existence of a remarkable optical excess (see e.g. Haberl 2005 and P{\'e}rez-Azor{\'{\i}}n et al. 2006).\\
The motion of the electrons is free parallel to the field lines but impeded by Larmor rotation perpendicular to them, the tensor component of the heat conductivity perpendicular to the field lines is strongly suppressed by the square of the magnetization parameter. Therefore, it is suggesting to consider the magnetic field to be the primary source for the observed anisotropies.\\
The thermal evolution of the crust is determined by the energy balance equation which has, in axial symmetry and with general relativistic effects included, the following form:
\begin{eqnarray}
\frac{e^{-\Lambda}}{r^2} \frac{\partial}{\partial r}
(r^2~F_r~e^{2\Phi}) + \frac{e^{2\Phi}}{r~\sin \theta}
\frac{\partial}{\partial \theta} (\sin \theta F_{\theta}) =
 \left(e^{\Phi} C_v \frac{dT}{dt} + e^{2\Phi} Q_\nu \right)\, \,,
\label{equ:heat-balance}
\end{eqnarray}
where $T$ is the local temperature, $e^{\Phi(r)},e^{\Lambda(r)}$ 
are the redshift and length correction factors, $F_r$ and $F_{\theta}$ 
are the local radial and meridional components of the heat flux and $r$ and $\theta$ the local coordinates.
$Q_{\nu}$ and $C_v$ are the neutrino emissivity and specific heat, respectively, per unit volume.
Studying stationary configurations and neglecting neutrino energy losses, the right-hand side of Eq.~\ref{equ:heat-balance}  can be set to zero, and it results in 
\begin{eqnarray}
\frac{1}{x^2} \frac{\partial}{\partial \tilde{x}}
(x^2~\tilde{F}_r) + \frac{1}{x~\sin \theta}
\frac{\partial}{\partial \theta} (\sin \theta \tilde{F}_{\theta}) = 0
\label{equ:heat-balance2}
\end{eqnarray}
where $x=r/R$,  $\partial/\partial\tilde{x} \equiv e^{-\Lambda} \partial/\partial x$ and
$\tilde{F}_{r,\theta} \equiv e^{2\Phi} F_{r,\theta}/R$. While in the envelope, the outer shell with densities $\rho \le 10^{10}$ g cm$^{-3}$, the magnetic field has both classical and quantum effects on the electron motion, in the crustal regions below the envelope, the quantized motion of electrons transverse to the magnetic field lines doesn't play any r{\^o}le for the magnetic modification of the  heat transport and the field acts dominantly via the classical Larmor rotation of the electrons. The components of the heat conductivity tensor $\hat{\kappa}$ and that of the temperature gradient determine the heat flux vector 

\begin{eqnarray}
e^{\Phi}\vec{F} =
- \hat{\kappa} \cdot \vec{\nabla} (e^{\Phi}T) =
- \; \frac{\kappa_0}{1+(\omega_{\scriptscriptstyle B}\tau)^2} \times
\nonumber \\
        \left[ \vec{\nabla} (e^{\Phi}T) +
       (\omega_{\scriptscriptstyle B}\tau)^2 \; \vec{b} \; (\vec{\nabla} (e^{\Phi}T) \cdot \vec{b}) +
			\omega_{\scriptscriptstyle B}\tau \; \vec{b} \times \vec{\nabla} (e^{\Phi}T)
        \right]\,\, 
\label{equ:Heat-Flux}.
\end{eqnarray}
For a prescribed magnetic field structure, which determines the components of the heat conductivity tensor, Eq.~\ref{equ:heat-balance2} is solved with the heat flux components given by Eq.~\ref{equ:Heat-Flux} until a stationary solution is found. With the temperature at the crust--core interface fixed, while the outer boundary condition uses the relation between the temperature at the bottom of the envelope and that at the surface derived by Potekhin \& Yakovlev (2001) takes all the complex physics of the heat flux through the strongly magnetized envelope into account, including the quantizing effects of the field on to the electron motion (for details see Geppert et al. 2004 and 2006).\\
The essence of the effect of a strong magnetic field is that the heat flux $\vec{F}$ is forced to be almost aligned with the local field $\vec{B}$ when 
$(\omega_{\scriptscriptstyle B}\tau)^2 \gg 1$ since then the component of the thermal conductivity tensor $\hat{\kappa}$ parallel to $\vec{B}$ is $\kappa_\| = \kappa_0$ while the components in the perpendicular directions are 
$\kappa_\perp = \kappa_0/(1+(\omega_{\scriptscriptstyle B}\tau)^2) \ll \kappa_\|$.\\
For a magnetic field configuration consisting of axially symmetric toroidal and poloidal costituents, the azimuthal component of the heat flux $F_{\varphi}$ is independent of $\varphi$ but certainly {\em not} equal to zero,
in spite of having $\partial T/\partial \varphi \equiv 0$. Since for strong fields heat essentially flows along the field lines, when
$\vec{B}^\mathrm{tor}$ is dominant, $F_{\varphi}$ will also be much larger than $F_{\theta}$ and $F_r$ and produce a winding of the heat flow around the symmetry axis: $\vec{F}$ follows the shortest possible paths with the highest possible conductivity and this winding effectively acts as a heat blanket.\\
Typical crustal temperature distributions with the corresponding surface temperature profiles are shown in ~\figref{fig:T_crust}. 
A noticeable general feature is the asymmetry between the two magnetic hemispheres, resulting from the asymmetry of the total field $\vec{B}$, since the dipolar poloidal crustal and star centered field constituents,
$\vec{B}^\mathrm{crust}$ and $\vec{B}^\mathrm{core}$, are anti-symmetric with respect to the equatorial plane, while the crustal toroidal field, $\vec{B}^\mathrm{tor}$, (as chosen here) is symmetric. In cases where the poloidal component is almost comparable to the toroidal one,  the asymmetry is barely detectable but in all other cases it is clearly visible. The star centered core field, which superimposes the poloidal and toroidal crustal fields causes practically no deviations from isothermality of the crust for densities $\rho > 10^{10}$ g cm$^{-3}$. In the envelope, however, that core field produces a meridional temperature gradient as shown by Greenstein \& Hartke (1983), recently refined by Potekhin \& Yakovlev (2001). If sufficiently strong, it may counteract the effects of the crustal field and tries to establish a temperature distribution closer to crustal isothermality. For a detailled discussion see Geppert et al. (2006).\\
The very distinct surface temperature distributions resulting from significantly non--isothermal crusts have several immediate obervational consequences. In presence of a strong toroidal field in the crust, the channeling of heat toward 
the polar regions results in the appearance of two hot spots of very reduced
size in comparison with the hot polar regions which would appear in case of an almost isothermal crust, having the magnetic field effects caused by a star centered field only. Figure~\ref{fig:Tdistr_Litc} shows five examples of surface temperature distributions and the resulting observable pulse profiles of the X--ray light curve. 
Naturally, models with the smallest hot spots result in the highest pulsed fractions, $P_f$, with values above 30\%, in contradistinction to the case of an almost isothermal crust which results in $P_f \sim 5$ \%.
The composite blackbody spectra resulting from the same five cases of Fig.~\ref{fig:Tdistr_Litc} are shown in Figure~\ref{fig:Spectra}. The distances to the model stars have been adjusted to give the same maximum flux in the X--ray band, and thus very similar X-ray spectra. Given this adjustment the differences between the relative areas of the hot and cold regions in the various cases result in differences in the predicted optical fluxes.
Comparison of the surface temperature plots (left panels of Figure~\ref{fig:Tdistr_Litc}) with the relative optical fluxes shows a direct correlation between the relative size of the cold region with the optical flux.
Obviously, the presence of two small warm regions separated by an extended cold belt has two immediate observational consequences. The first one is that the observable pulsed fraction in the X-ray band can be very large, above 30 \% assuming isotropic blackbody emission. As the second one, the emission of th cold region contributes little to the X-ray flux but dominates the detectable flux in the optical range, appearing as an ``optical excess''. These successes in explaining the observations are strong indicators, that the heat transfer through the strongly magnetized crust of isolated NSs is indeed responsible for the small hot spots.\\
However, the above discussed dipolar axisymmetric field configurations produce symmetric, but not sinusoidal, light--curves. For RBS 1223 and RX J0720.4-3125 the light--curves are clearly not symmetric. This could be interpreted by assuming that the hot spots are not in antipodal position but have a meridional distance of $\approx 160^\circ$ (Schwope et al. 2005, Haberl et al. 2006, Haberl 2006). The non--uniqueness of the light curve interpretation allows also an axisymmetric arrangement of hot regions. A superposition of dipolar and quadrupolar magnetic field constituents could as well be able to produce precise fits of the observed pulse profiles of the ``Magnificent Seven'' (Zane \& Turolla 2006).\\
Indeed, very recently, P{\'e}rez-Azor{\'{\i}}n et al. (2006a) have shown that a crustal field configuration consisting of dipolar and quadrupolar parts in both the toroidal and the poloidal constituents produces a warm equatorial belt in addition to the polar hot spots. The corresponding surface temperature distribution explains convincingly well all observational evidences seen for RX J0720.4-3125, a prominent member of the ``Magnificent Seven'', namely the X--ray spectrum, the ``optical excess'', the pulsed fraction, the spectral feature around 0.3 keV, and the light curves including their strong anti--correlation of the hardness ratio with the pulse profiles in both the hard and the soft band. It is compelling that these model calculations rely on an axisymmetric field configuration instead of a non--axisymmetric one which could also explain the light curve by the non--alignement of the northern and southern hot spot. The latter model, however, implies a complicated structure of the currents which maintain the non--axisymmetric field. Since it is hard to believe that such a field can be stable over a long period, the model of  P{\'e}rez-Azor{\'{\i}}n et al. (2006a) is perhaps likely to be realized in isolated NSs.\\

\section{Discussion and Conclusion}\label{sec:5}
The magnetic field of NSs is a complex entity, maintained by currents which circulate both in the core and in the crust. While the former support essentially the large scale, long living ($\gtrsim 10^8$ yrs) dipolar field which is responsible for the rotational evolution, the latter have a considerably shorter decay time ($\sim 10^{6\ldots 7}$ yrs) and cause the anisotropic heat tranport through the crust, its cracking, and the Joule heating. The crustal field may consist of a toroidal and poloidal part. The large scale modes of the latter join at the surface the star centered core field.  For the typical pulsar lifetime the core and crust field "collaborate" to establish the conditions for radio emission. Thus, a strong sub--surface toroidal field could provide via Hall--drift induced processes the small scaled field structures necessary to produce sufficient electron positron pairs in the polar gap. Sometimes it is argued by means of population synthesis results that the NS magnetic field decays - if at all - on timescales which exceed the typical pulsar lifetime ($\sim 10^7$ yrs) considerably. The population syntheses, however, reflect almost only the effect of the core field on the rotational evolution, which is affected by the poloidal part of the crust field only during its shorter lifetime.\\  
At the here discussed turning points the NS's magnetic field will evolve into qualitatively different ways. Should the inborn field be stabilized against MHD instabilities, the NS has a perspective as magnetar, otherwise it becomes a "standard" radio pulsar. Depending on the power of fallback accretion and on the electric conductivity of the crust, the NS will appear as a radio pulsar soon after its creation in a supernova or will evolve with a weak surface field which has minor braking effects on the rotation. If temperature gradients in the crust are strong enough and maintained for a sufficient long period, a magnetic field may be rapidly generated and the NS becomes a pulsar in spite of heavy fallback accretion. Exceeds the magnetization parameter significantly unity locally and/or temporally both the magnetic and thermal evolution will proceed differently from that of a weakly magnetized NS. This may have observational consequences both for the rotational and cooling history.\\
Although the basic ideas of the here discussed physical processes are known, there is still a lot of work necessary to understand them in more detail. This concerns both the properties of NS matter (e.g. its conductivity) and the processes around the NS's birth (initial $\alpha$, $P$, and field configuration) as well as the nonlinear and non--axial symmetric processes of field evolution. Since the  NS's life is so intimately connected with the magnetic field, any better insight into its evolution will return a better understanding of the physics of the most fascinating stellar objects in the universe.\\

\section{Acknowledgment}
I gratefully acknowledge collaboration and discussions with W. Becker, F. Haberl, D. Page, J. Pons, K.-H. R{\"a}dler, M. Rheinhardt, and J. Tr{\"u}mper. I am especially grateful to J. Pons and M. Rheinhardt for carefully reading this manuscript.
%
%

%
%



\printindex
\end{document}